\documentclass[11pt,letterpaper]{article}
\usepackage{arxiv}

\usepackage[pdftex]{graphicx}
\usepackage{geometry}
\usepackage{rotate}
\usepackage{wrapfig}
\usepackage{caption}
\usepackage{subfigure}
\usepackage{times}
\usepackage{amsmath,amstext,amsbsy,amsopn,amsthm,amsfonts,amssymb}
\usepackage{url}
\usepackage{lastpage}
\usepackage{algorithm}
\usepackage{balance}
\usepackage{placeins}
\usepackage{subfigure}
\usepackage{caption}
\usepackage{capt-of}
\usepackage{algorithmic}
\usepackage{fancyhdr}
\usepackage{wrapfig}
\providecommand{\keywords}[1]{\textbf{\textit{Keywords---}} #1}
\begin{document}

\title{iGLU 4.0: A Continuous Glucose Monitoring and Balancing Paradigm with Physiological Parameters}

\author{
	\begin{tabular}{ccc}
		Prateek Jain & Amit M. Joshi & Saraju P. Mohanty \\
		Electronics \& Instr. Eng. & Electronics \& Commu. Eng. & Computer Science and Engineering   \\
		Nirma University, India & 	MNIT, Jaipur, India. & University of North Texas, USA. \\
		prateek.jain@nirmauni.ac.in &amjoshi.ece@mnit.ac.in & saraju.mohanty@unt.edu
	\end{tabular}	
}

\maketitle

\cfoot{Page -- \thepage-of-\pageref{LastPage}}
\begin{abstract}
The conventional method of glucose measurement such as pricking blood from the body is prevalent which brings pain and trauma. Invasive methods of measurement sometimes raise the risk of blood infection to the patient. Sometimes, some of the physiological parameters such as body temperature and systolic blood pressure (SBP) are responsible for blood glucose level fluctuations. Moreover, diabetes for a long duration usually becomes a critical issue. In such situation, patients need to consult diabetologist frequently, which is not possible in normal life. Therefore, it is required to develop non-invasive glucose balancing paradigm, which measures blood glucose without pricking blood along with physiological parameters measurement and decision model. The proposed paradigm helps to doctor, who is even available at remote location. There won't be any need to consult frequently. In the way of optimized non-invasive system design, an NIRS technique with specific wavelengths along with physiological parameters is taken to predict the precise glucose value. The all parameters (glucose, Blood pressure and body temperature), food intake and insulin levels are parts of decision model, which would help to the doctor to take decision related to the further medicine doses and diet plan. The patients would have suggestions according to maintain their blood glucose level. The proposed system demonstrated an accurate model with MARD and AvgE 12.50\% and 12.10\% respectively using DNN model. Coefficient of determination $R^2$ has been found 0.97.
\end{abstract}

\keywords{Smart Healthcare, Healthcare Cyber-Physical System (H-CPS), Non-invasive Glucose Measurement, Glucometer}
%
\section{Introduction}
\begin{figure}[htbp]
	\centering
	\includegraphics[width=0.5\textwidth]{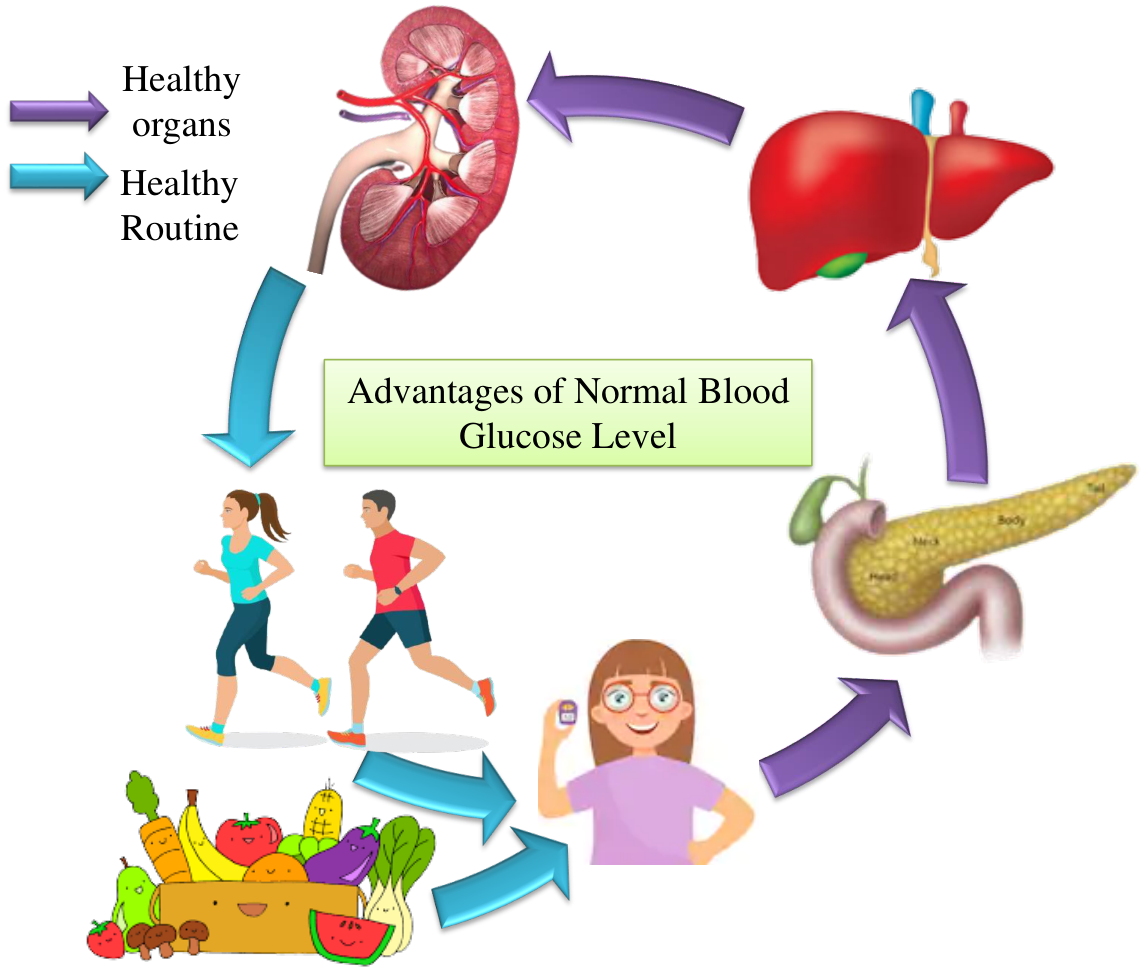}
	\caption{Advantages of normal blood glucose level}
	\label{balanced_routine}
\end{figure}
Now a days, balancing of blood glucose has become one of the global challenges for human life. To keep blood glucose normal, people are preferring balanced routine and prescribed diet plan. There are several advantages of an ideal blood glucose level. Because of ideal blood glucose levels, maximum internal organs will be healthy. Due to this, a person will be able to keep a healthy routine for a quality life. The advantages of normal blood glucose level is represented in Fig.\ref{balanced_routine}.

When the person has an issue of balancing the body glucose level then, it refers to Glycemic imbalance. The imbalance in context of hyperglycemia is the prime factor in increasing the probability of being a diabetic patient. An unbalanced diet is one of the main factors for the occurrence of diabetes Mellitus. Hence, timely diagnosis and better treatment of diabetes are focusing points of research of recent healthcare trends. Therefore, it is required to analyze the factors which are responsible for fluctuating the blood glucose. There are lots of points that claim for unexpected variations of glucose levels in the human body. In common cases, four points have been considered for variation in blood glucose levels which is represented in Fig. \ref{factors}.
Many cases have been observed for blood glucose fluctuation because of changes in physiological parameters. Person use to report to the medical consultant with several issues. particular medical expert use to check the diagnostic reports of all testing for particular diagnosis as fluctuation in any parameters may be possible because of other's deficiencies. Hence, it is required to verify, validate and provide correct treatment for the particular cause as per the reliable and expert healthcare. Hence, it is required to analyze the other body parameters, which are responsible to fluctuate the blood glucose level in the case of diabetic patients. For this objective, the physiological parameters are required to compute along with blood glucose level values to provide an accuracy using advanced device.
It has been analyzed that blood glucose variations are not possible due to change in prescribed diet plans and workouts instantly. opposite of this, continuing fever in the body or hypertension may be the factors for the sudden change in blood glucose level. Hence, various physiological parameters and other factors are required to compute along with blood glucose level values for training of the advanced device. With the advancement of further work, patient can be able to control the blood glucose with more precision. By using such type of paradigm, the consultant would be able to justify the current issues and provide instant treatment without any hurdles. Such paradigm provides the transparency in current situation and precise diagnosis. By using such technology, some more constraints can be overcome in the era of blood glucose balancing. It has been observed that the type of insulin doses are specified in terms of short and long acting for different cases. As per the advancement in healthcare technology, it is required to provide the necessary details to the medical consultant belonging to the patients at same time even if users are at remote locations.  
 
The overall paper has been arranged in the given sequence. Significance of physiological parameters are demonstrated in Section \ref{Sec:parameters}. Prior research work is reported in Section \ref{Sec:Prior-Works}. Novel contribution of present work is elaborated in Section \ref{Sec:Novel}. Proposed system is described with process of glucose detection in Section \ref{Sec:Proposed-Work}. Section \ref{Sec:Calibration} elaborates machine learning models for system calibration and validation. The experimental analysis has been done in Section \ref{Sec:error analysis}.
\section{Why do we need to consider physiological parameters}
\label{Sec:parameters}

\begin{figure}[htbp]
	\centering
	\includegraphics[width=0.5\textwidth]{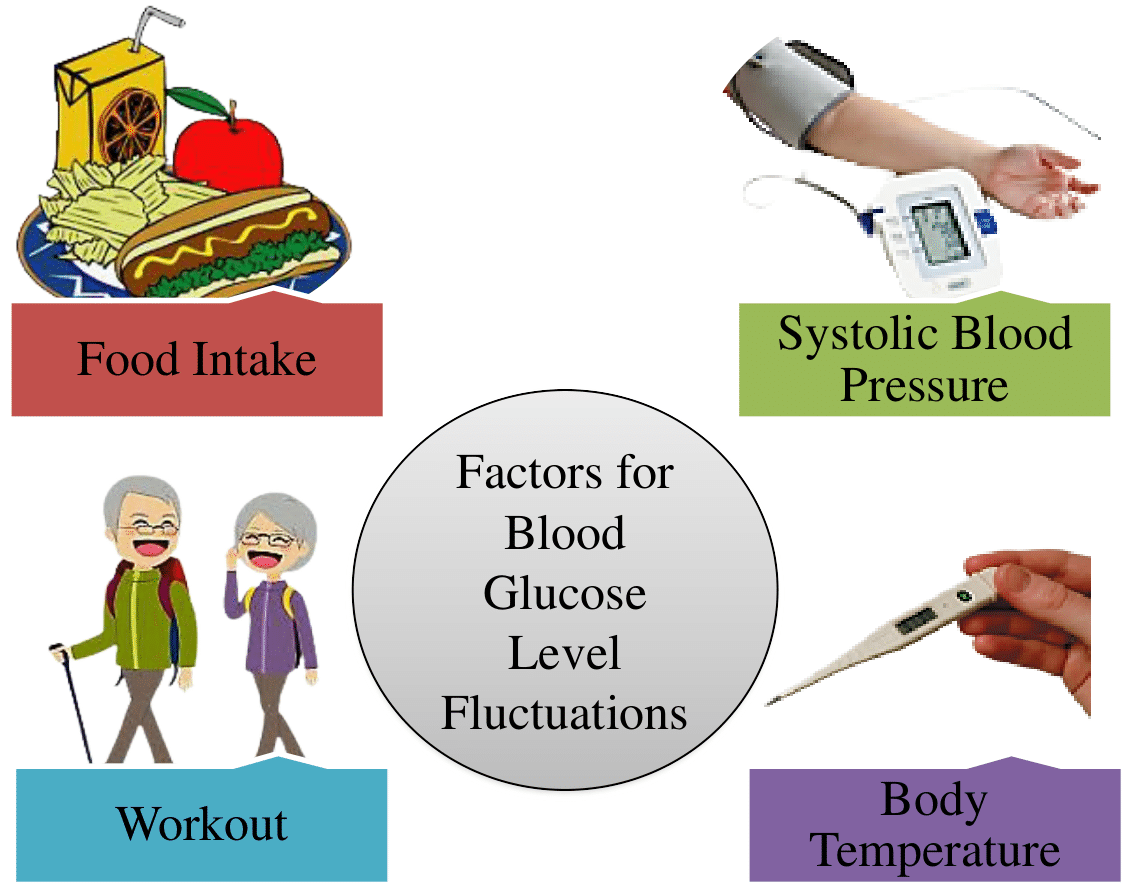}
	\caption{Representation of factors affecting blood glucose level}
	\label{factors}
\end{figure}

Diabetes is one of the highly challenging diseases for the human health in the world. The WHO also declares that 79\%-82\% of diabetic deaths occur in developing countries \cite{Pai2017}. For diabetes care, it is necessary to design the blood glucose monitoring system which could be available in rural and urban area with minimum cost and ease to use \cite{Song2015}. Presently, the most of successive and commercially available blood glucose measuring devices are invasive. These devices are accurate and recommended to blood glucose test for health care. Highly accurate conventional laboratory test has been preferred for diabetes patients \cite{Pai2018}. These type of blood glucose measurements are not better for frequent monitoring. For diabetic patients, it is necessary to measure multiple times in a day, which causes irritation and trauma. At the other end, the body blood is contaminated because of consumable lancets and strips. To avoid these problems, non-invasive glucose monitoring paradigm has been explored in previous research work \cite{jain2019precise}. Sometimes, it has been observed that the blood glucose level fluctuates with respect to the imbalance of physiological parameters. Hence, the genuine prediction of body glucose is only possible, when physiological parameters could be predicted at the same time. Then, precise diagnosis would be possible. Prior non-invasive systems have been designed without consideration of physiological parameters and diabetic history of patients. Therefore, consultant won't be able to provide a precise treatment in specific cases such as patients with fever or hypertension along with corresponding history of medicine intake. To overcome these issues, a novel non-invasive system with history of medicine intake is required to propose with integration of physiological parameters such as SBP and body temperature for real-time validation. These two parameters are responsible for variation in blood glucose which is shown in Fig.\ref{factors}.
The proposed system is calibrated according to blood glucose readings in fasting, postprandial and random testing modes. Then, system is tested through comparison of non-invasive and corresponding invasive predicted blood glucose values.

\section{Prior Related Research Work}
\label{Sec:Prior-Works}
various approved rapid blood glucose measurement products are dependent upon a cost-effective electrochemical mechanism. They use to measure blood glucose with high accuracy without prior arrangements. Usually, pricking the blood is the initial stage of the whole process for glucose measurement. Measurements with less duration are so much irritating due to pricking for samples more than 3-4 times in a day for frequent monitoring. The minimally invasive approach based implantable biosensors have been proposed earlier for continuous measurements with glucose oxidase generation. These are implanted beneath the skin for continuous measurement. The photometric technique was also attempted to detect glucose molecule in blood drops.

The minimally invasive approaches were used to develop frequent glucose monitoring sensors. The wearable sensors are implanted to analyze the results for continuous glucose monitoring from the membrane which contains the immobilized glucose oxidase. Glucose monitoring biosensors are configured with auxiliary frameworks to provide the environment of continuous glucose monitoring through implantation mechanism. These minimally invasive methods have experimented on intensive care required diabetic patients. The wearable sensors based micro system was explored for continuous glucose monitoring to make the portable solution for frequent monitoring. Likewise, there was a solution for continuous monitoring using a biosensor with the help of a transponder chip. The response obtained from the transponder chip was processed for the calibration for the semi-invasive technique of the Dexcom sensor. Diabetes balancing has been explored through sensors with the system nominated as an artificial pancreas system. The minimally invasive techniques based designs are identified with some sort of limitations such as precision in glucose values and will have a shorter life span for monitoring \cite{9415429}. 

Optical methods based blood glucose measurement systems have been proposed to detect the blood glucose without pricking the blood from fingertip. Various bio-potential based techniques have been explored to detect the blood glucose of human being \cite{Yilmaz2014}. Using these techniques, the detection of glucose molecule through the skin resistance, dielectric constant and capacitance has been proposed \cite{Liu2016}. Glucowise is a non-invasive blood glucose monitoring device which is currently under test for clinical purpose \cite{Chaplin}. This device explored the impedance spectroscopy technique. The change in the blood glucose concentration depends upon the impedance variation. However, the change in object position may be refer to blood glucose variation. Hence, the device will not be precise for all patients. A lot of work have been proposed for glucose detection through human tears, saliva and sweat substance. Presently, the continuous wave multi-wavelength photo-acoustic technique also discussed for glucose detection \cite{Tajima2017}. Pai et.al. proposed the pulse width modulated photo-acoustic spectroscopy technique for glucose detection \cite{Pai2018}. However, there is not yet commercialized viable solution for medication purpose. After consideration of further advancement and benefits of NIR spectroscopy, an initial prototype setup has been calibrated and validated for precise blood glucose measurement. This prior work was a motivation to develop an edge computing intelligent device for same application. Then, iGLU an intelligent blood glucose measurement system has been developed for smart healthcare \cite{Jain_IEEE-MCE_2020-Jan_iGLU1}. This prior intelligent system was able to measure capillary blood glucose with almost 98\% accuracy. Further, iGLU 2.0 has been developed with same methodology to measure serum glucose measurement \cite{iGLU2}. That proposed system reported massive performance parameters in terms of predicting genuine blood glucose level as serum glucose is usually considered the exact level for proper medication. Following to this work, iGLU 1.1 is proposed with glucose-insulin control virtual paradigm, which reported an analytical studies on various standard subjects of diabetic patients \cite{9221132}. IGLU 3.0 was developed to provide a specific feature of device and data security. An optimized secure iGLU is formed using standard encrypted algorithm \cite{iGLU3}. Many non-invasive systems are proposed to provide precise glucose measurement system. All the methods till date are suitable for real time blood glucose measurement only. It is required to fulfill the objective gaps in terms of cause of blood glucose fluctuation. Hence, it is required to predict the other factors which are directly and indirectly responsible for blood glucose fluctuation. Here, the most important factors such as body temperature and blood pressure are highlighted to justify blood glucose fluctuation.

On basis of literature work and spectrum analysis, it is also analyzed that glucose detection will be more accurate involving optical detection method \cite{jain2019precise}. Therefore, 940 and 1300 nm wavelength specific light is used for glucose detection with prediction of physiological parameters in current work.\\
The body temperature and systolic blood pressure are key parameters which may be related to the blood glucose concentration. If body temperature rises, then arteries walls dilate which enhance for insulin consumption. This may result in low blood glucose level. Besides of this, consistently high blood glucose may result in hardening of arteries walls. This is one of the causes of being high blood pressure. Table \ref{prior_work} demonstrated the prior work done in non-invasive paradigm along with specific features advancement in current work. 

\begin{table*}[htbp]
	\caption{Non-invasive Works in Technology and Features Perspective}
	\label{prior_work}
	\centering
	\begin{tabular}{llllll}
		\hline
		\textbf{Works}&Method&Technology&Cost&Reliability&Functional\\
		\textbf{}&&&&&Specification\\
		\hline		 \hline
		Song, et al. \cite{Song2015}&IMPS+NIRS&Sensing+&Moderate&Moderate&Glucose\\
		&&Measurement&&&Measurement\\
		\hline
		Pai, et al.	\cite{Pai2018}&Photo-acoustic&Sensing+&High&High&Glucose\\
		&&Measurement&&&Measurement\\
		\hline
		Jain, et al.	\cite{jain2019precise}&mNIRS&Sensing&Low&Moderate&Glucose\\
		&&&&&Measurement\\
		\hline
		Jain, et al.	\cite{Jain_IEEE-MCE_2020-Jan_iGLU1}&mNIRS&Sensing &Low&High&Glucose\\
		&&+Trained Model&&&Measurement\\
		\hline
		Joshi, et al.	\cite{iGLU2}&mNIRS&Sensing &Low&High&Serum Glucose\\
		&&+Trained Model&&&Measurement\\
		\hline
		Joshi, et al. \cite{iGLU3}&mNIRS+&Sensing &Low&High&Body Glucose, Insulin\\
		&Data Security&+Trained Model&&&Measurement\\
		\hline
		Jain, et al.\cite{9221132}&Mathematical&Trained&Moderate&Moderate&Glucose\\
		&Model&Model&&&Insulin Model\\
		\hline
			Murad, et al.\cite{new9431682}&mNIRS&Simulation&-&-&Glucose Detection\\
			\hline
			Kirubakaran, et al.\cite{newkirubakaran2023antiallergic}&Microwave&Sensing&High&Moderate&Glucose\\
			&&&&&Measurement\\
			\hline
			Mohammadi, et al.\cite{newmohammadi2023dual}&Microwave&Sensing&High&Moderate&Glucose\\
			&Resonator&&&&Detection\\
		\hline
		\textbf{(iGLU 4.0)}&\textbf{mNIRS+}&\textbf{Sensing+}&\textbf{Moderate}&\textbf{High}&\textbf{Glucose Balancing}\\
		&\textbf{Model}&\textbf{Measurement}&&&\textbf{Paradigm}\\
		\hline
	\end{tabular}
\end{table*}

\section{Novel Contribution of Proposed Work}
\label{Sec:Novel}

The proposed blood glucose balancing paradigm is painless which provides the exact cause of blood glucose fluctuations in terms of hyperglycemia. It also reduces the probability of being blood related diseases as the measurement is possible without pricking the blood. Proposed optical approach along with optimized prediction model for estimation and physiological parameters prediction have resolved the issue of precise and rapid diagnosis with cause and treatment. The proposed system has no need to prior setup for measurement. The system can be used to measure the blood glucose of any person anytime. The system will be portable after packaging to use everywhere. The system is easy to use, fast operated and low cost solution for smart healthcare.
Proposed iGLU 4.0 is used to provide the facility of continuous monitoring as well as other vital parameters estimation, which are responsible for fluctuation. The proposed intelligent device is used to direct the consultant for reliable treatment to the patients. The features advancement is demonstrated in Fig. \ref{Sec:Prior-Works}. 

\begin{figure}[hbp]
	\centering
	\includegraphics[width=1.0\textwidth]{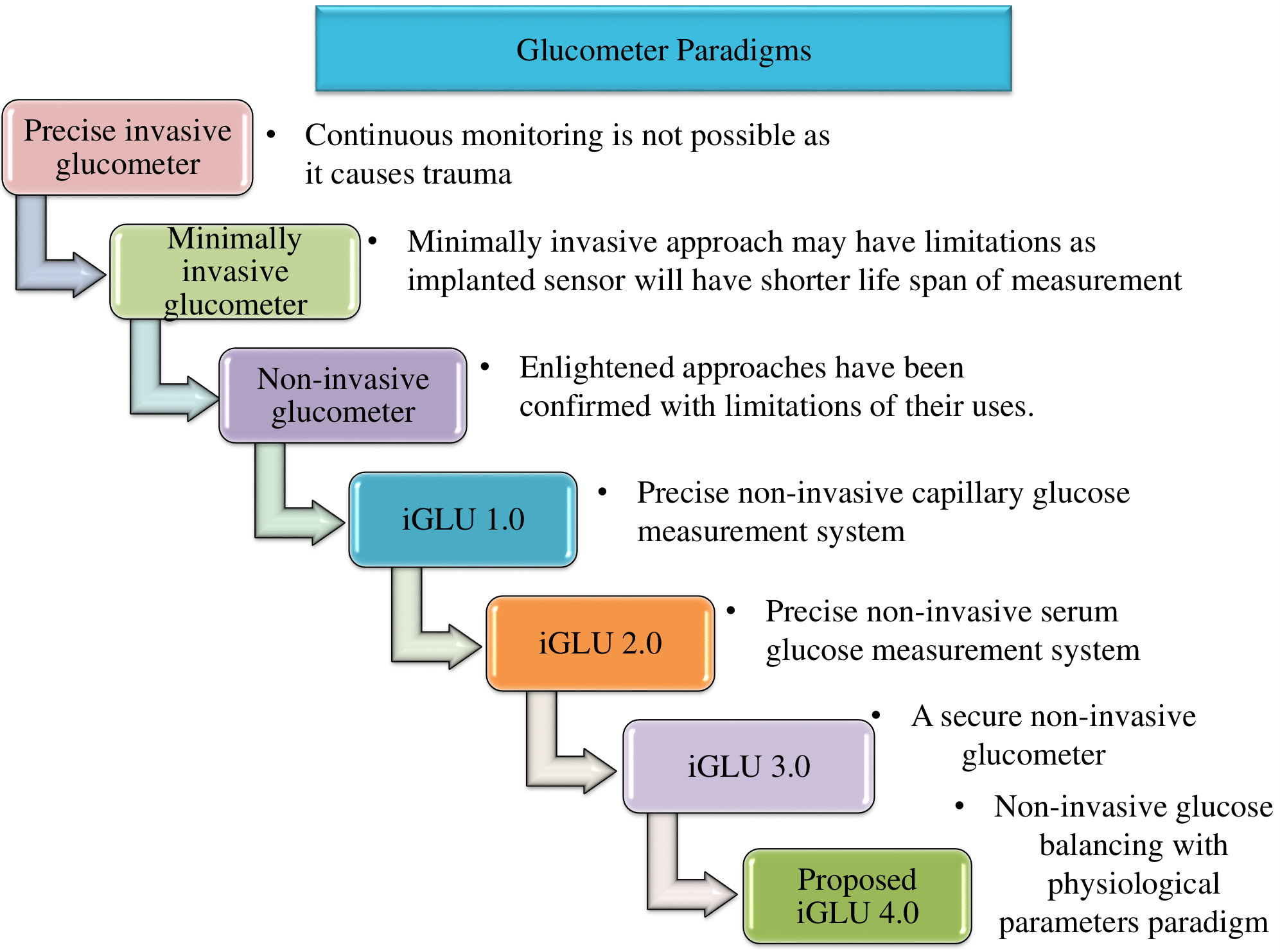}
	\caption{Representation of glucose measurement paradigm with challenges}
	\label{prior work}
\end{figure}
The \textbf{novel contribution in proposed work} are the following: 
\begin{enumerate}
	\item 
	An accurate non-invasive glucometer is developed with prediction of other physiological parameters. The continuous monitoring of parameters are logged  and accessible by users using IoMT framework.
	\item 	The optimized prediction model has been calibrated, validated and tested using appropriate data-set for precise diagnosis of blood glucose fluctuations.
	\item The blood glucose values are analyzed with respect to the other reports of each category of person, which demonstrate the reliable paradigm for consultant located at remote location to provide better treatment. 
 
\end{enumerate}

\begin{figure}[htbp]
	\centering
	\includegraphics[width=1.0\textwidth]{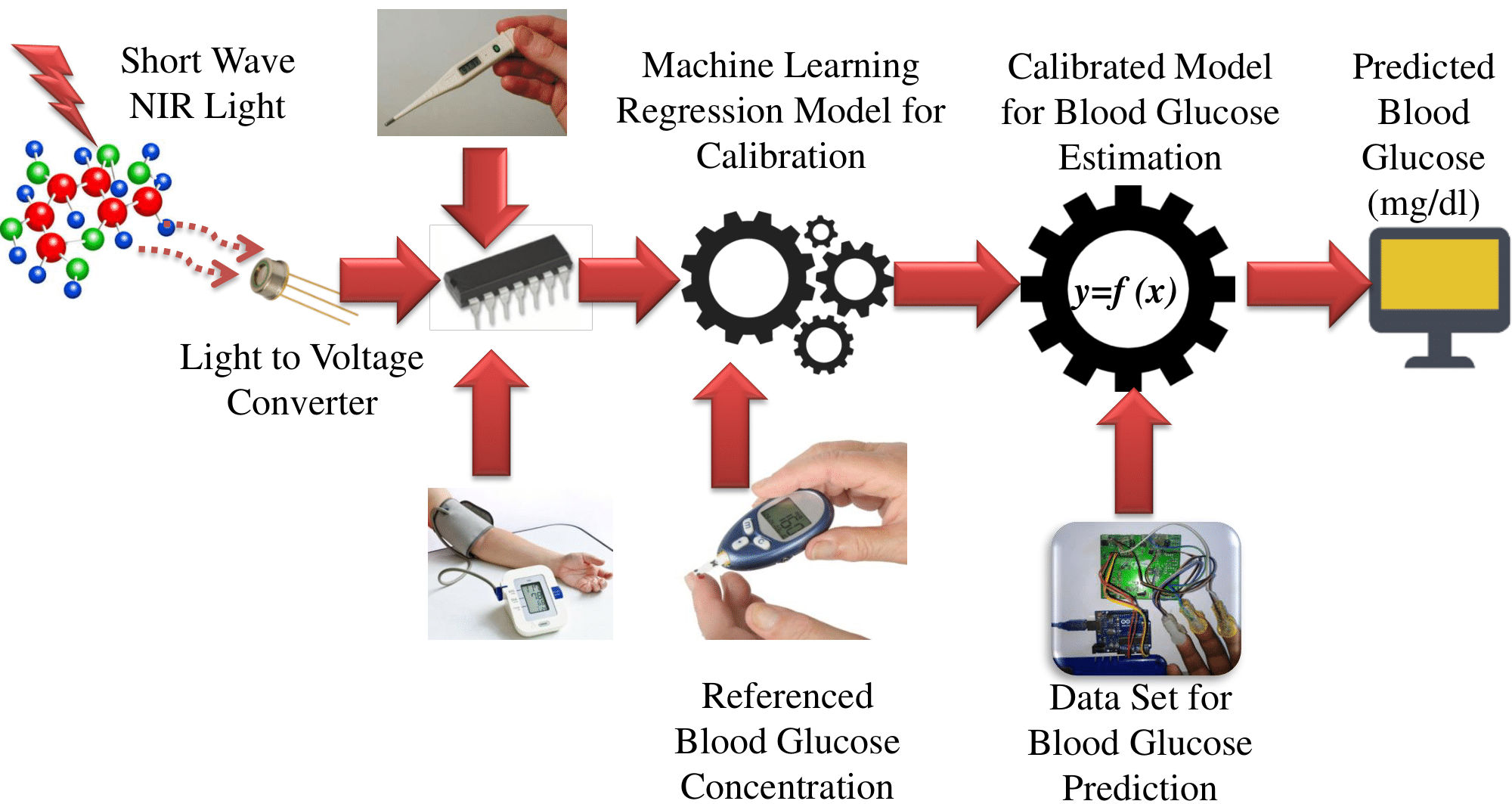}
	\caption{The pictorial view of process flow for proposed glucometer.}
	\label{theme}
\end{figure}

\section{Proposed Non-invasive Glucose Balancing paradigm with Physiological Parameters}
\label{Sec:Proposed-Work}

The prediction of blood glucose has been done by placing the fingertip between the gap of transmitter and receiver sensor.


The different configuration of sensors generate the responses in voltages. These voltage values are processed along with systolic blood pressure and body temperature for estimation of predicted blood glucose through data acquisition module. Proposed system design of non-invasive glucose detection consists of optimized three channel multi-wave reflection and absorption spectroscopy.\cite{Jain_IEEE-MCE_2020-Jan_iGLU1}. 
The pictorial view of process flow for proposed glucometer is represented in Fig. \ref{theme}.
The estimated glucose values are considered from an optimized decision model, which analyze the data of scheduled food intake and insulin secretion in the body. Based on the decision data, medical consultant would be able to diagnose the actual blood glucose values and cause of fluctuation, if it occurs. 

\begin{figure}[htbp]
	\centering
	\includegraphics[width=1.0\textwidth]{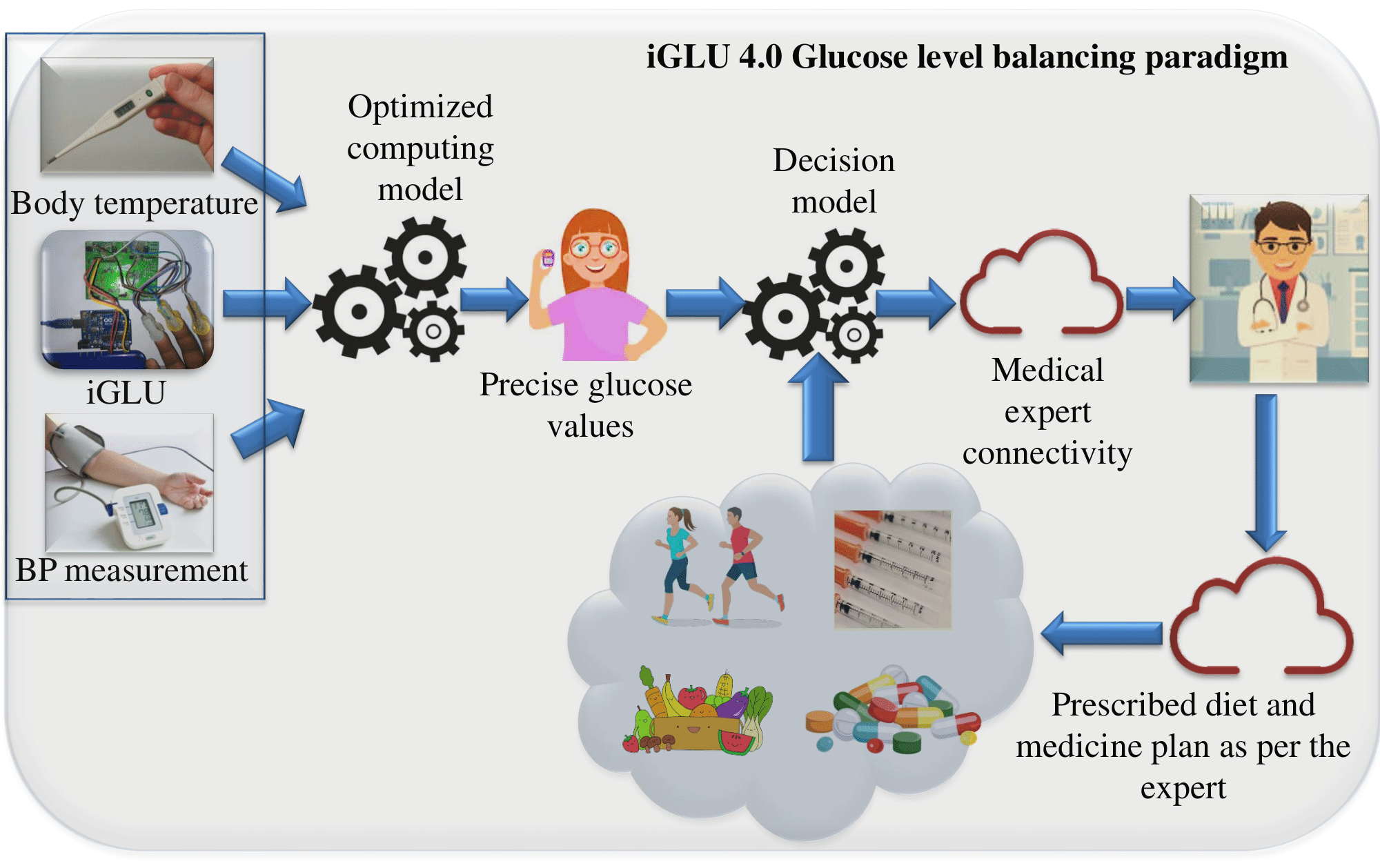}
	\caption{Representation of proposed non-invasive glucose sensing paradigm}
	\label{architecture}
\end{figure}

According to the Fig. \ref{architecture}, the precise glucose values are obtained after estimation of physiological parameters. The proposed system is used to balance the glucose values after analyzing the data of other parameters. The main significance of the proposed system is that the medical consultant would have prior information about patients and their related things. Medical consultant needs to provide treatment based on prior given information, which is provided by proposed system. This feature will reduce enough effort and time consumption to justify the cause of fluctuation and further treatment will be provided earlier.

\section{Proposed Machine Learning based Regression Model with Validation}
\label{Sec:Calibration}

Post processing computation model is used to interpret the response of proposed system in estimated glucose value for monitoring. It is necessary to analyze the post processing model for precise estimation of output \cite{Sarangi2014}. The characteristics of collected data are represented in Table \ref{dataset}.

\begin{table}[htbp]
	\caption{Characteristics of Collected Data}
	\label{dataset}
	\centering
	\begin{tabular}{lll}
		\hline
		\hline
		\textbf{Samples Basic} & \textbf{Samples with} & \textbf{Samples with}\\
		\textbf{Characteristics} & \textbf{Glucose Values} & \textbf{Physiological Parameters}\\
		\hline
		\textbf{Gender} & \textbf{Hyperglycemia} & \textbf{SBP}\\
		Male:-   \textbf{62} & Male:-   \textbf{40} & All samples with\\
		Female:- \textbf{54} & Female:- \textbf{30} & different SBP\\
		\hline
		\textbf{Age (Years)} & \textbf{Hypoglycemia} & \textbf{Body Temperature}\\
		Male:-   \textbf{22-65} & Male:-   \textbf{08}& All samples with\\
		Female:- \textbf{17-43} & Female:- \textbf{10} & different body\\
		&&temperature\\
		 & \textbf{Healthy} &\\
		   & Male:-   \textbf{14} &\\
		   & Female:- \textbf{14} &\\
		\hline
		\hline
	\end{tabular}
\end{table}

\begin{figure}[htbp]
	\centering
	\includegraphics[width=1.0\textwidth]{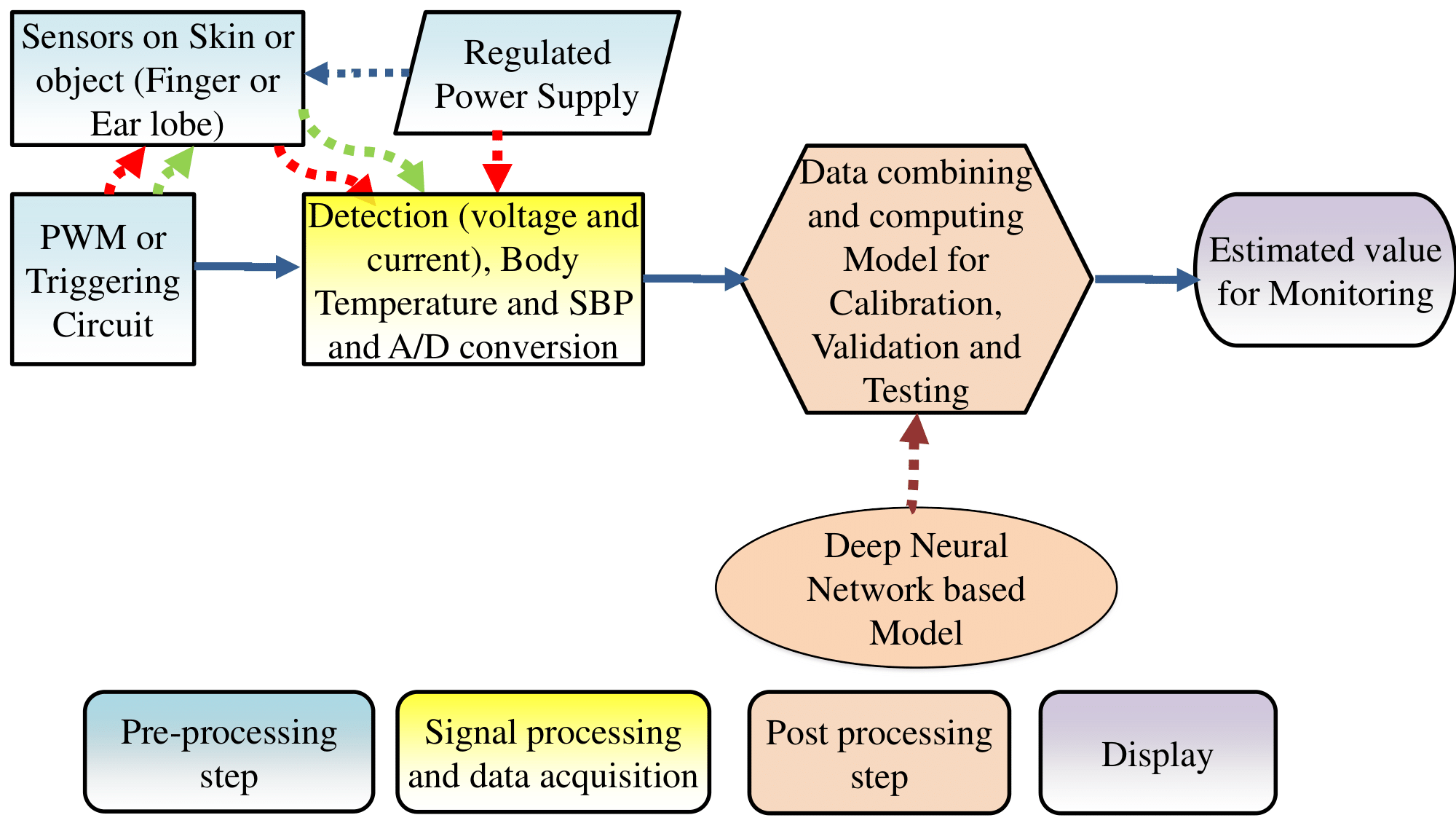}
	\caption{Processing steps of proposed glucose measurement}
	\label{Schematic}
\end{figure}

The diagram of processing steps of proposed glucose measurement system is given in Fig. \ref{Schematic}.

\begin{figure}[htbp]
	\centering
	\includegraphics[width=1.0\textwidth]{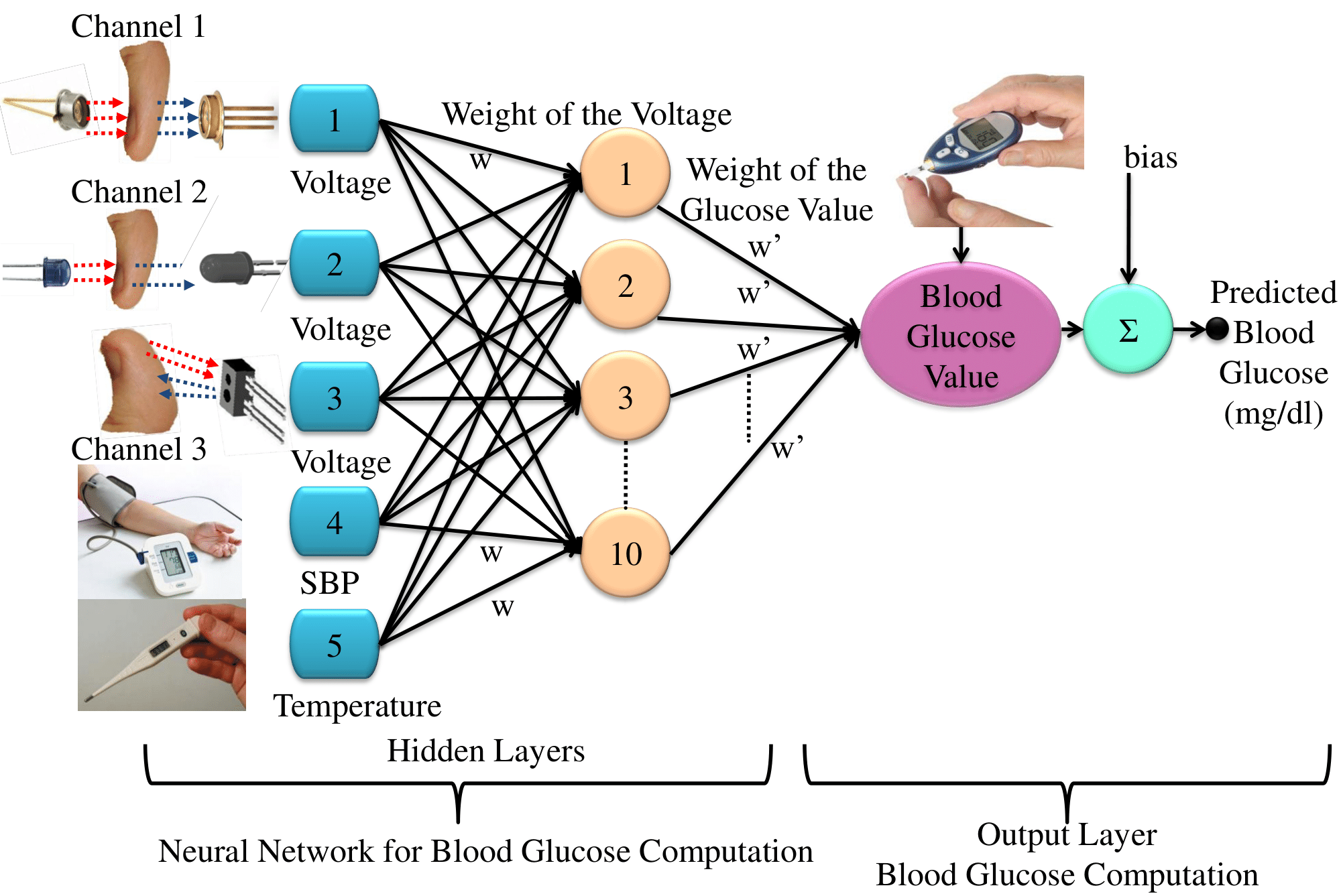}
	\caption{Block diagram of proposed DNN based model}
	\label{nnfit}
\end{figure}

DNN based model is also explored for statistical data analysis, which is generally used to simulate the model. The block diagram of proposed DNN model is given in Fig. \ref{nnfit}. It formulates the complicated relationship between attributes. Here in this proposed model, measured data of five channels are combined through DNN based model \cite{Song2015}.

The explored neural network model combines three inputs from multi-wave NIR spectroscopy and physiological parameters for the precise estimation of blood glucose \cite{Ming2009}. The feed forward back propagation neural network fitting model with sigmoid hidden neurons is used. The total nodes in hidden layers is 10. The network model is trained with levenberg-marquardt algorithm \cite{Song2015}. The sample size is 116, from which 34 subjects have been chosen randomly. DNN model is trained by randomly selected 24 subjects. The model is validated and tested through 10 subjects as per the standard ratio of model. After applying proposed DNN based fitting model, 12.50\% mARD and 12.10\% average error were found in predicted blood glucose values. The responses of predicted blood glucose value using DNN based model are represented in Fig. \ref{wave}. These samples have been taken from people of age group 17 to 65. The data has been collected from people in fasting, postprandial and random modes.
The ordinary mathematical model is used to provide the decision data to the consultant. The model is constructed with the independent variables such as food intake, insulin data and glucose values. The dependent variables are other parameters, which are used to analyze the glucose fluctuations.
The glucose level fluctuation as per the time is represented by the Eq. \ref{1} \cite{9221132}.

\begin{equation}
\frac{dG_{pls}}{dt}=\frac{[G_{gutabs}(t)+NHGB(t)-G_{iig}(t)-G_{renalg}(t)]}{V_{gd}}
\label{1}
\end{equation}

Here, $G_{gutabs}(t)$ = glucose absorption from the gut,
$G_{iig}(t)$ = without insulin glucose,  $G_{renalg}(t)$ = Renal glucose value and $V_{gd}$ = distributed glucose.\\

The total glucose of plasma is representing the glucose consumption and insulin concentration \cite{9221132}.

\begin{equation}
G_{iig}(t)=\frac{G_{pls}(c.EI(t)+G_{pls}G_{guii})(k+G_{ref})}{G_{ref}(k+G_{pls})}
\label{10}
\end{equation}

c= Slope between oral glucose (carbs and food intake) and insulin level, $G_{guii}$ = glucose consumption without insulin affect, $G_{ref}$= Reference glucose values and [$EI(t)$]= Effective insulin.

The glucose level in body will have source from food in terms of carbs intake which is demonstrated as below

\begin{equation}
\Delta G_{main}=G_{Emp}-(K_{gabsorb}\times G_{main})
\label{11}
\end{equation}

Here, $G_{gut}$= The body glucose and $G_{Emp}$= Gastric emptying.\\
$K_{gabsorb}\times G_{main}$= Glucose consumption for systemic circulation.\\
The above represented equations are used to estimate different parameters, which demonstrate the cause of blood glucose fluctuations. These equations analyze the proper cause and helps to provide treatment in terms of prescribed diet, medicines and insulin secretion plans so that patients could be better in balancing of glucose levels within short duration. 

\section{Analysis of Experimental Results}
\label{Sec:error analysis}

\begin{figure}[htbp]
	\centering
	\includegraphics[width=0.5\textwidth]{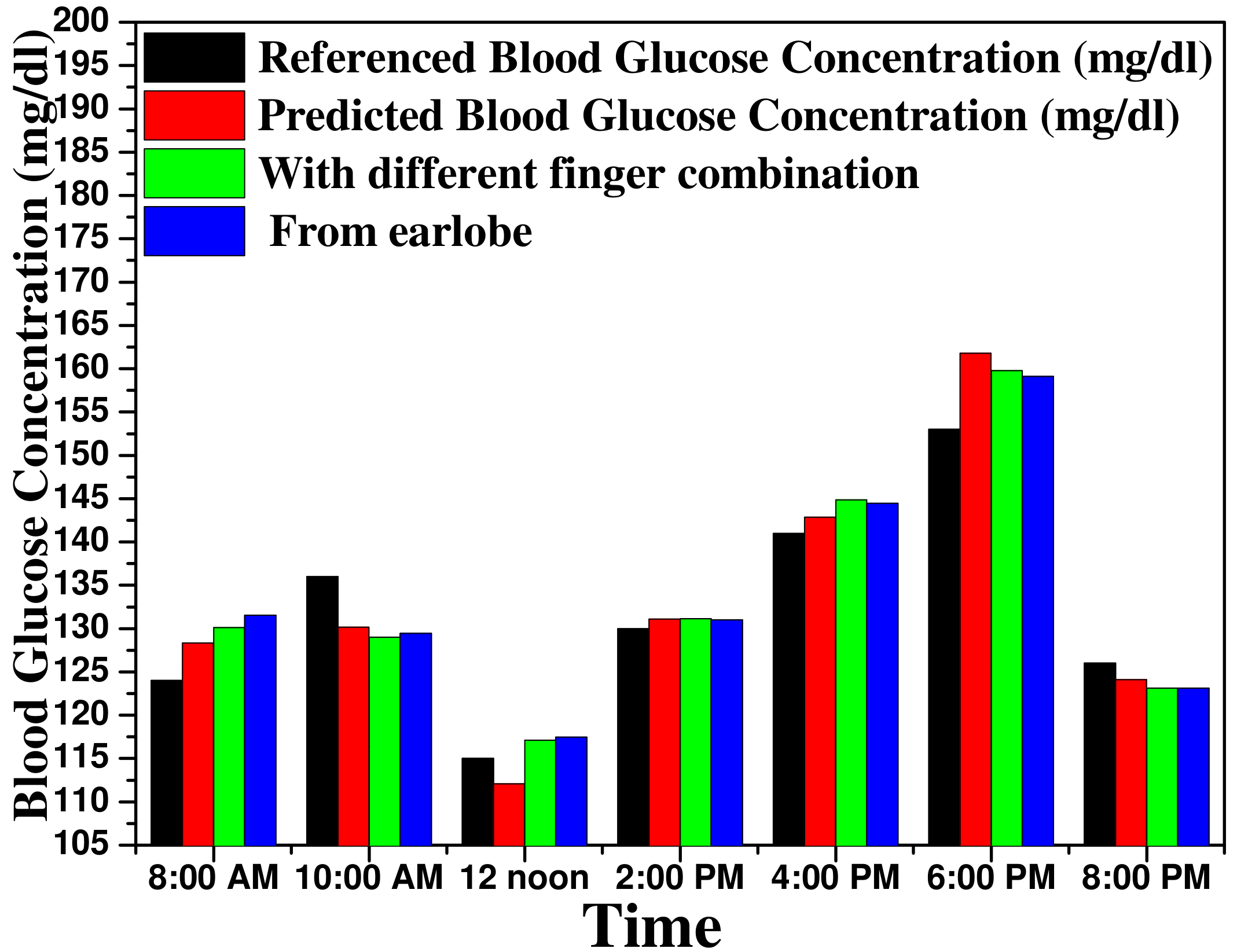}
	\caption{Predicted and referenced blood glucose concentration values through different objects}
	\label{ana}
\end{figure}

\begin{figure}[!h]
	\centering
	\subfigure[with body temperatures]{\includegraphics[width=0.45\textwidth]{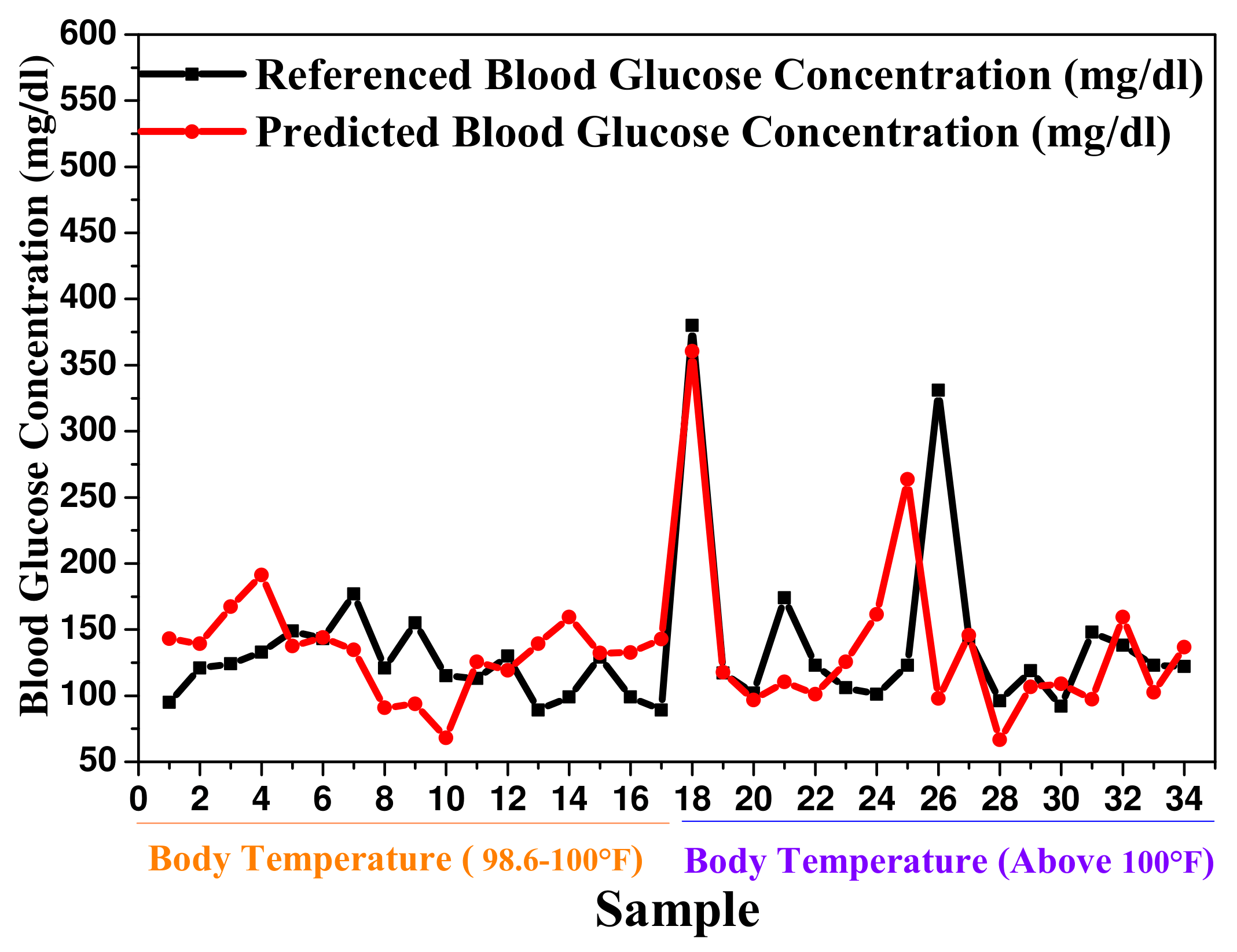}}\label{}
	\subfigure[with systolic blood pressures] {\includegraphics[width=0.45\textwidth]{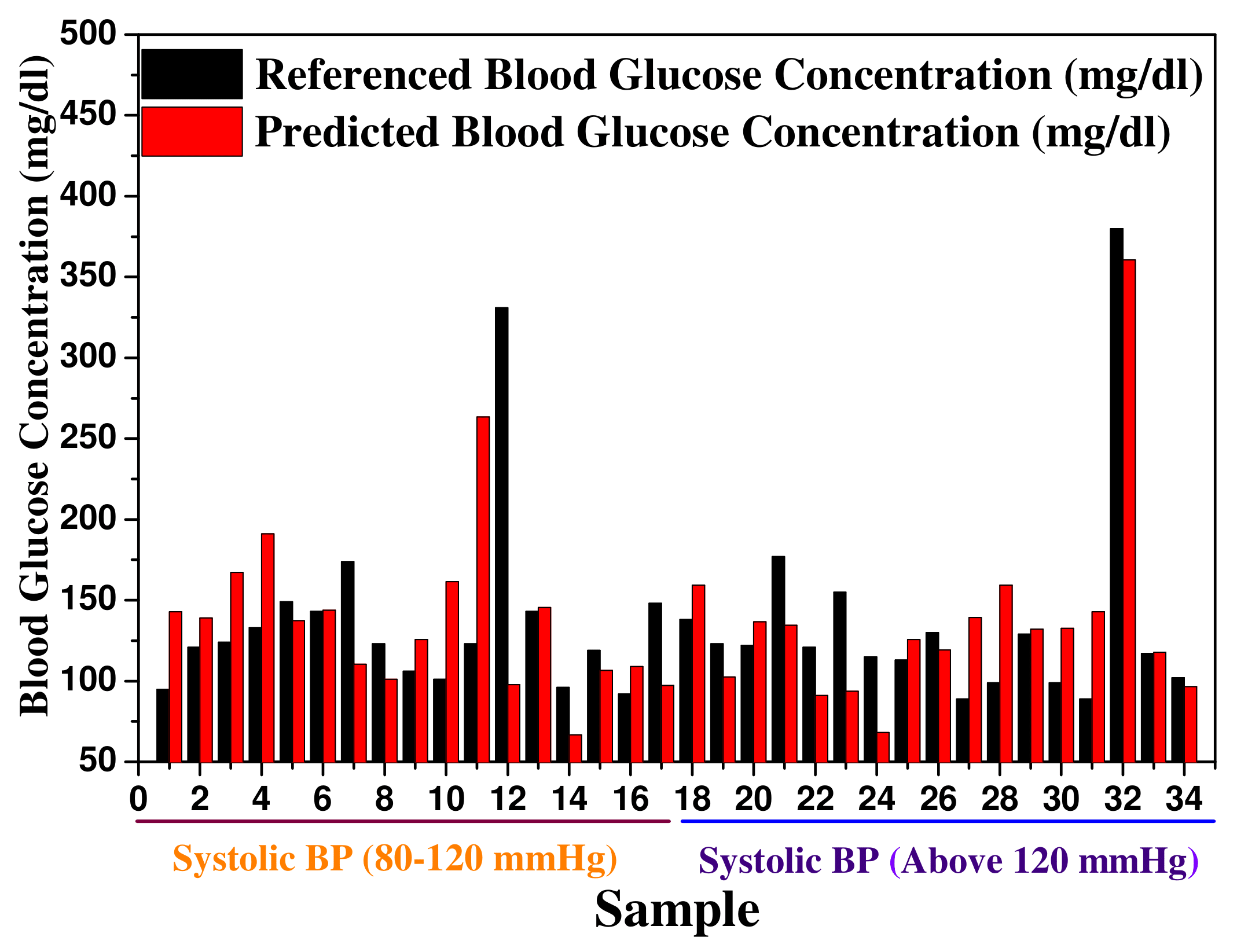}}\label{}
	\caption{Predicted and referenced blood glucose concentration} 
	\label{wave}
\end{figure}

Glucose measurement without pricking blood through optical detection is an optimized approach, which enhance the accuracy level at desired level. The finger size and different boneless part of the body doesn't affect the blood glucose concentration. To examine this, an experimental analysis is performed which represents that S/N ratio is not affected with the change of path length. A person aged 30 has been included for analysis. The blood glucose measurement has been done seven times with time intervals of 2 hours in a day. At the same time, the data has been collected from different combinations of fingers and earlobes. During this experimental work, it has been analyzed that there is not any impact of object change and finger and other boneless part. The referenced and predicted blood glucose values through different objects are shown in Fig. \ref{ana}.
This was an experimental analysis to validate the glucometer for precise measurement.
In next part of experimental analysis for validation of proposed glucometer, these 34 samples are randomly taken among 116 samples of male and female with their body temperatures and systolic blood pressure values. The graphical representations validate the accuracy of proposed glucometer with consideration of physiological effects. The predicted and referenced blood glucose values with physiological parameters are shown in Fig. \ref{wave}.

\begin{figure}[htbp]
	\centering
	\includegraphics[width=0.5\textwidth]{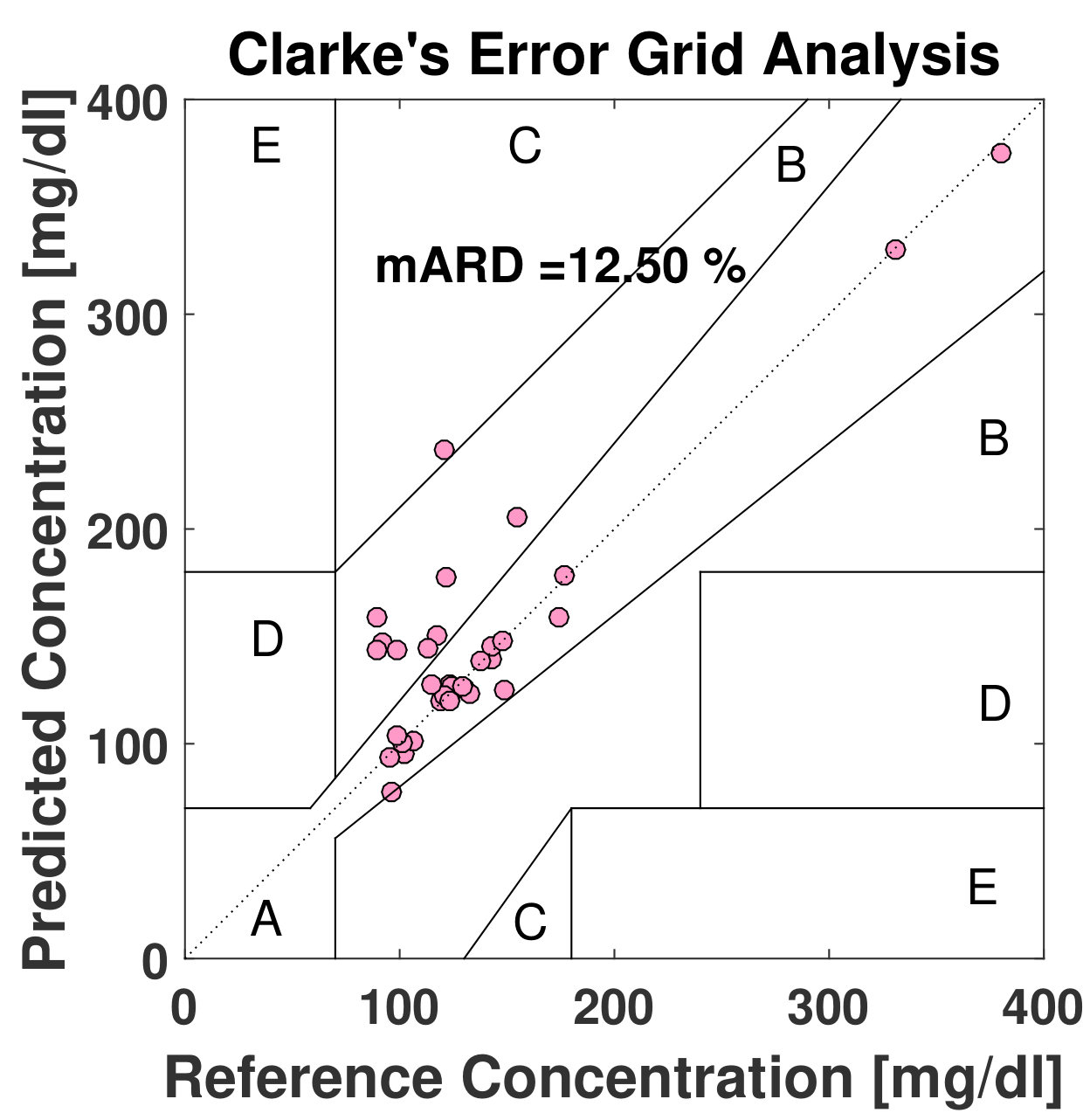}
	\caption{Predicted and referenced blood glucose concentration using deep neural network based model} 
	\label{cl1}
\end{figure}

To quantify the clinical accuracy of estimated blood glucose from proposed system, it is necessary to compare measured values from proposed system to the reference values from conventional method of blood glucose measurement. Clarke has elaborated the values which exists in zone A and B are desirable \cite{Clarke1987}. During clarke error grid analysis, it has been found that 94.12\% predicted values of blood glucose of samples exist in zone A and zone B using proposed system. The clarke error grid analysis of predicted blood glucose values are represented in Fig. \ref{cl1}.

\begin{figure}[htbp]
	\centering
	\subfigure[]{\includegraphics[width=0.5\textwidth]{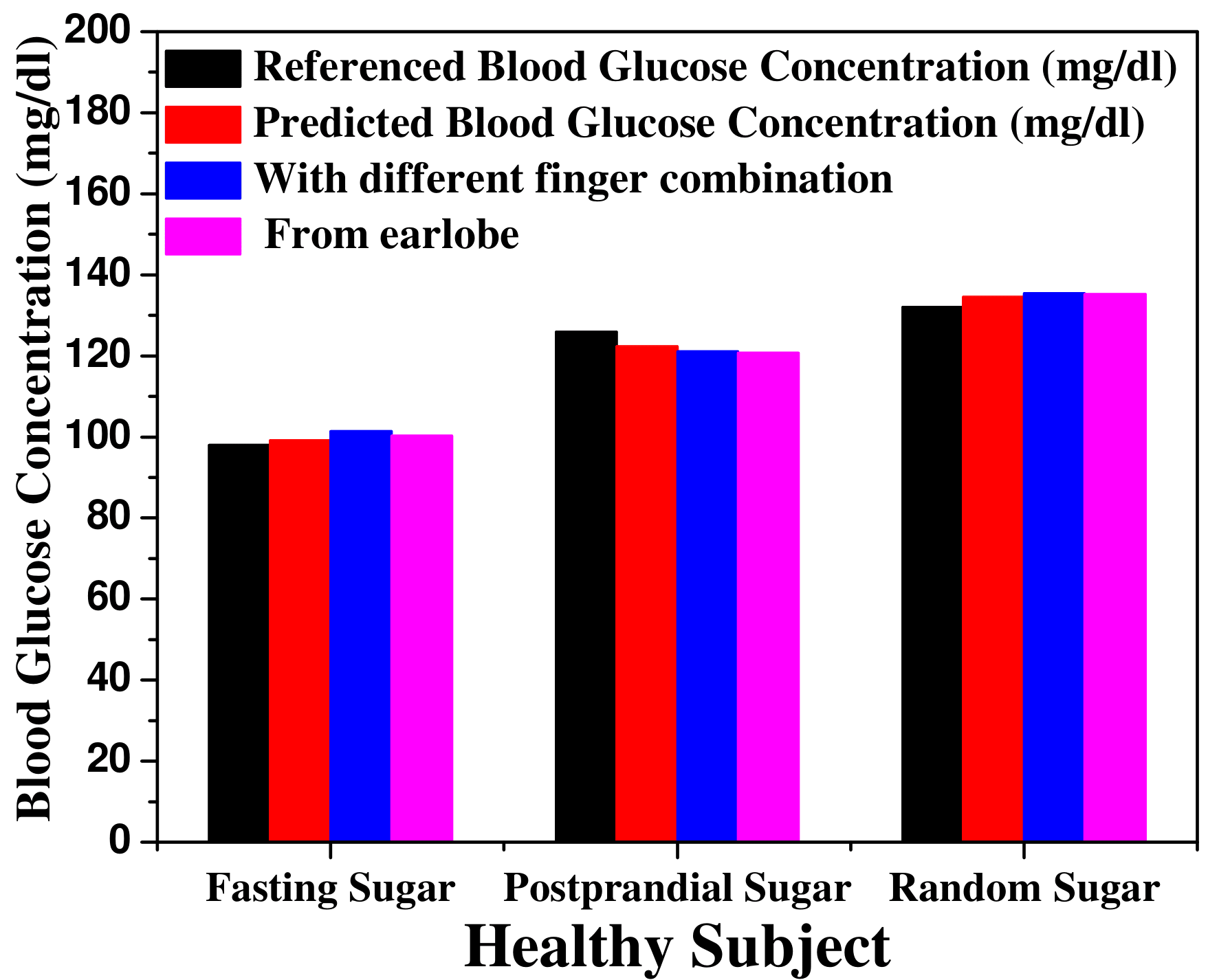}}\label{}
	\subfigure[] {\includegraphics[width=0.5\textwidth]{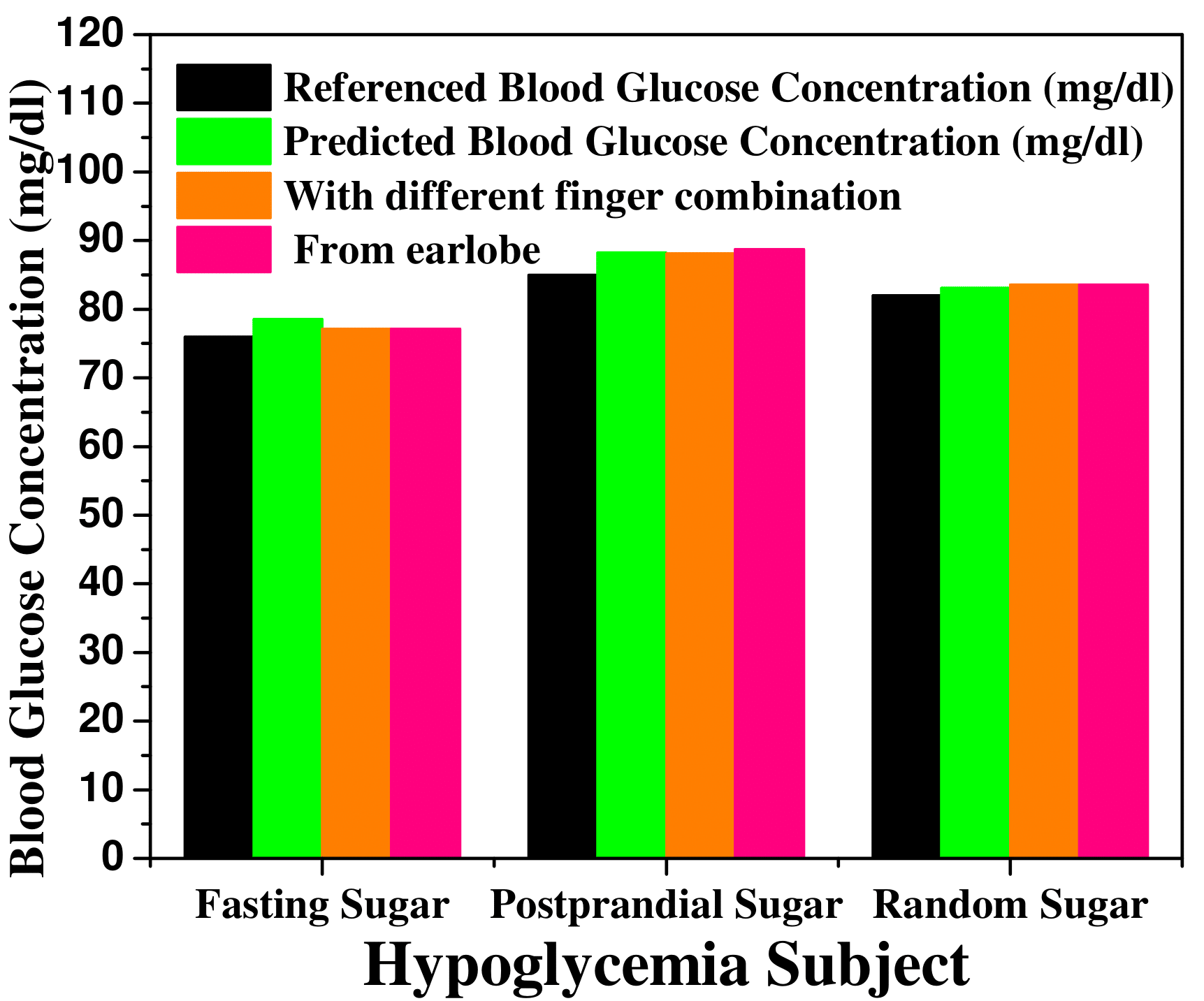}}\label{}
	\subfigure[]{\includegraphics[width=0.5\textwidth]{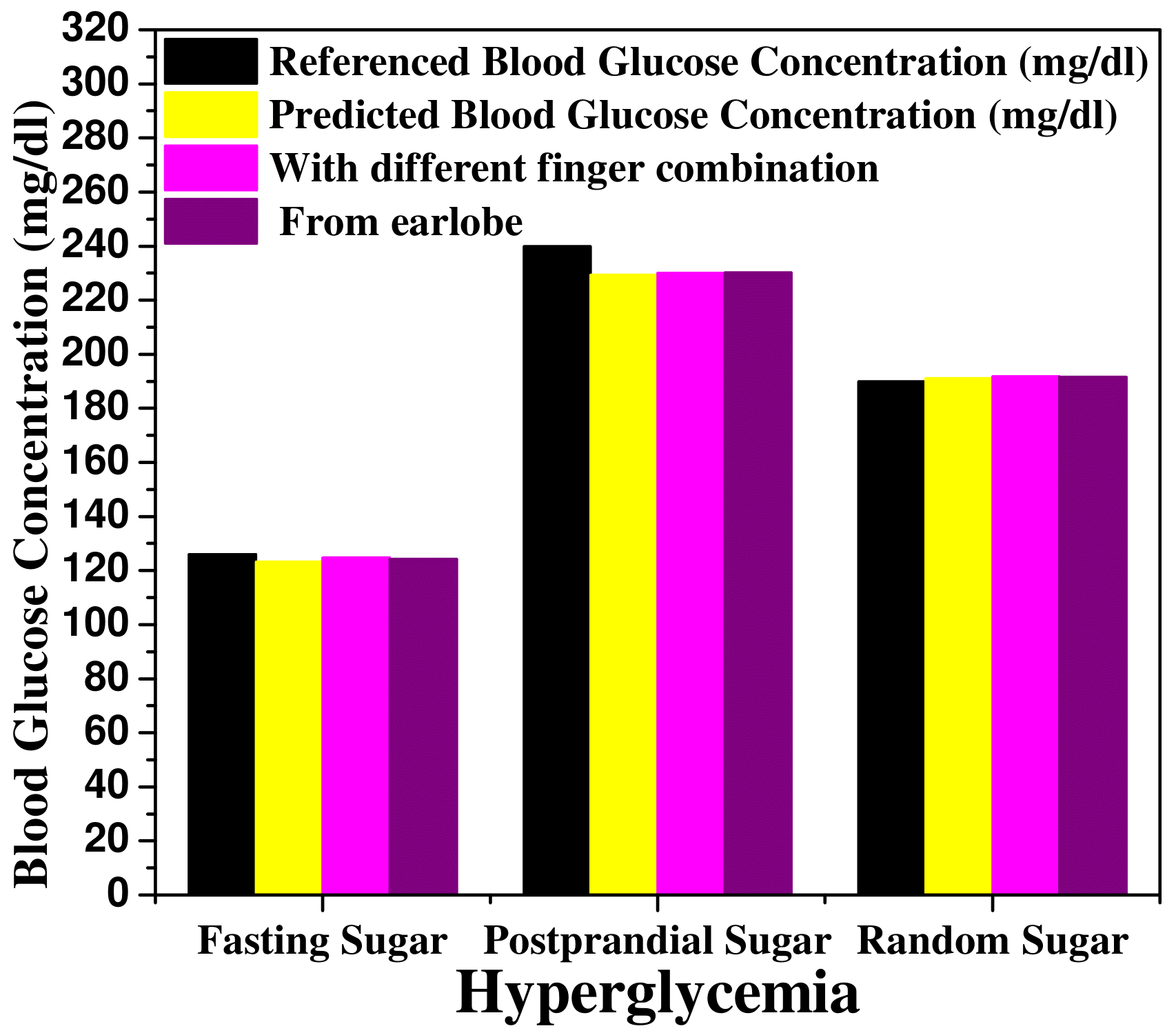}}\label{}
	\caption{Representation of system stability and result validity} 
	\label{stability}
\end{figure}

To analyze the device stability and results validity, three volunteers have been randomly selected, in which each belongs to the normal, hypoglycemic and hyperglycemic conditions. The precision level of proposed device is represented in Fig.\ref{stability}.
These different types of experimental analysis were compulsorily required to validate the proposed glucometer for precise measurement with physiological parameters considerations. 

\begin{figure}[htbp]
	\centering
	\subfigure[]{\includegraphics[width=0.45\textwidth]{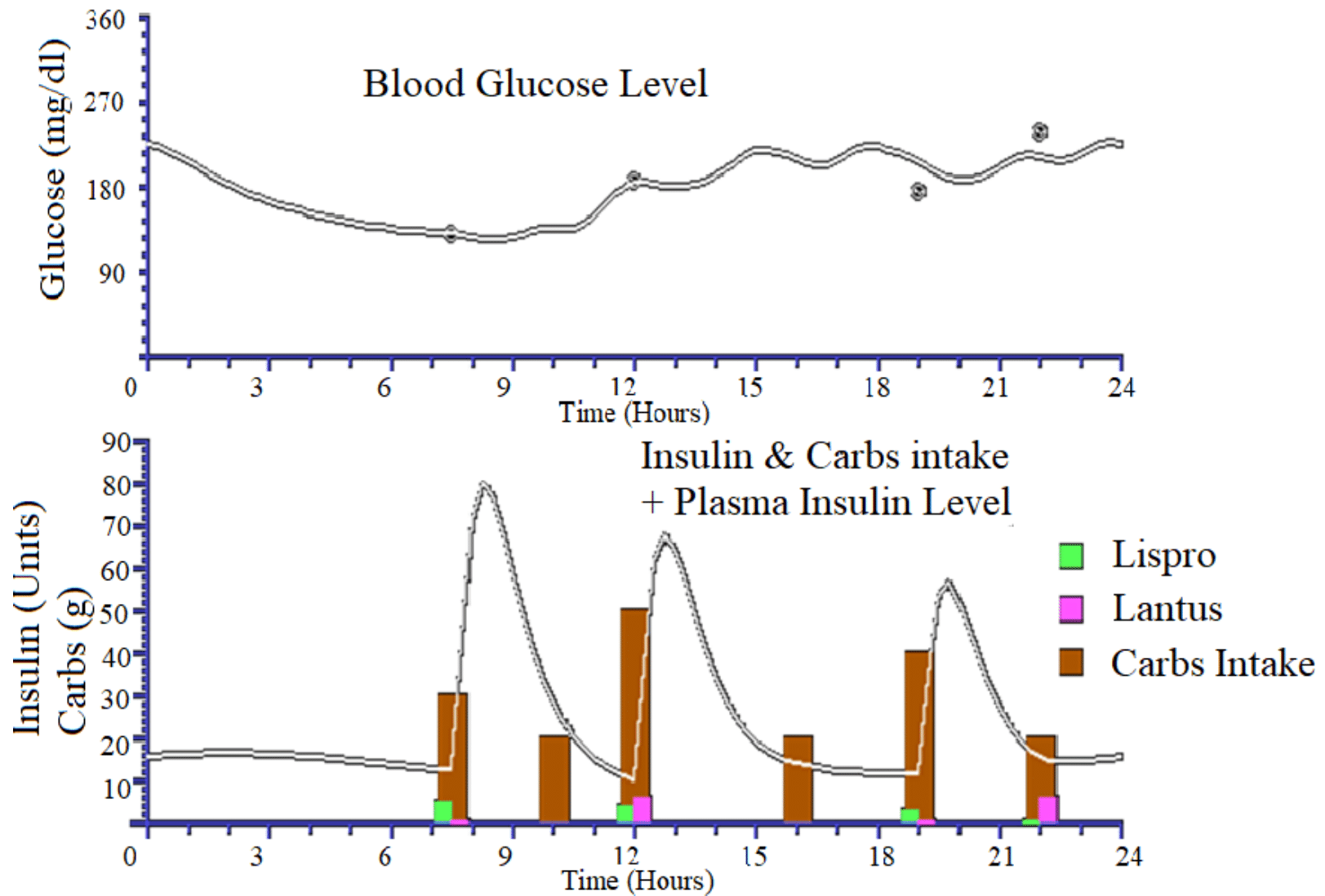}}\label{}
	\subfigure[] {\includegraphics[width=0.45\textwidth]{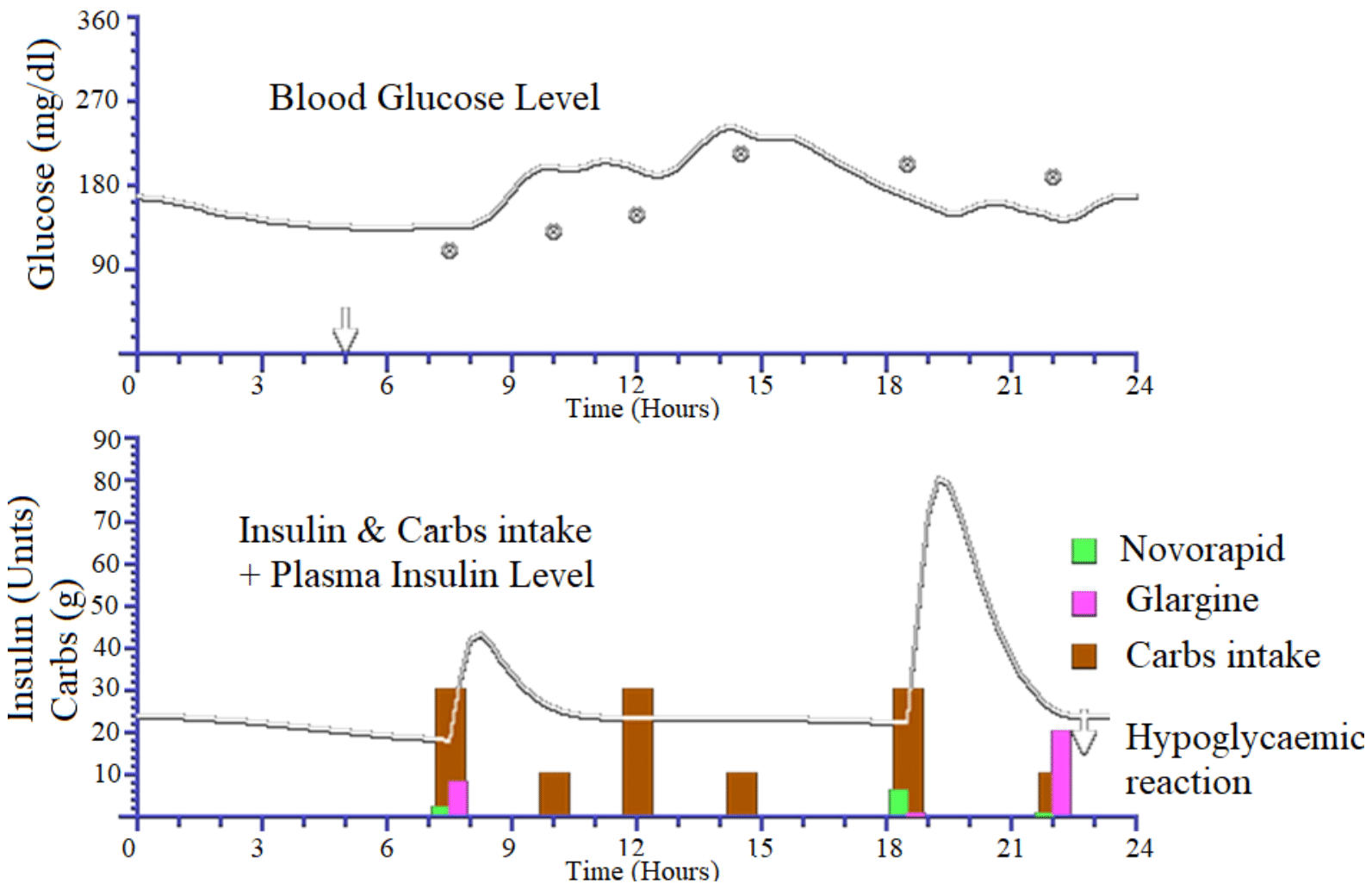}}\label{}
	\subfigure[]{\includegraphics[width=0.45\textwidth]{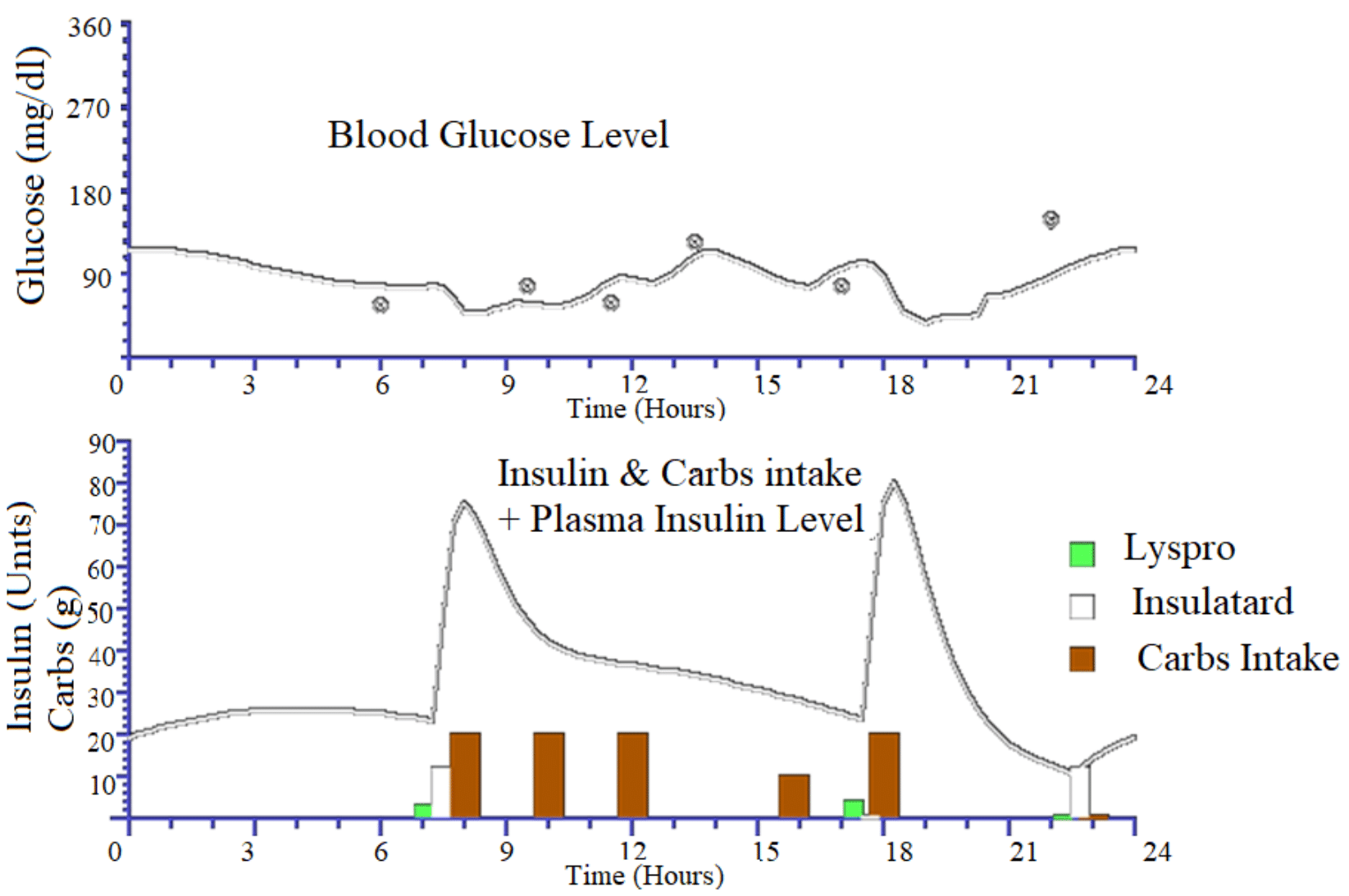}}\label{}
	\subfigure[]{\includegraphics[width=0.45\textwidth]{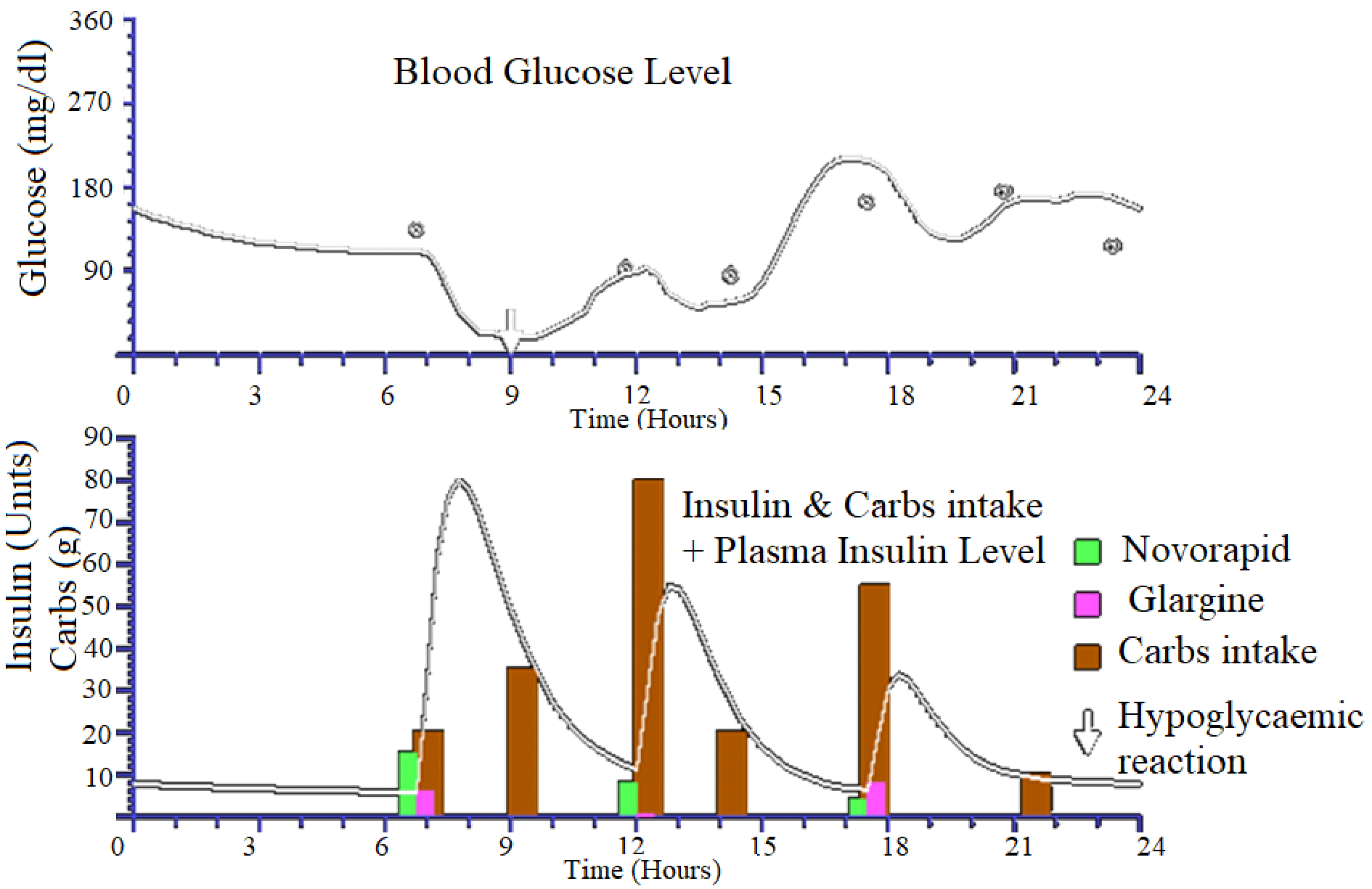}}\label{}
	\subfigure[]{\includegraphics[width=0.45\textwidth]{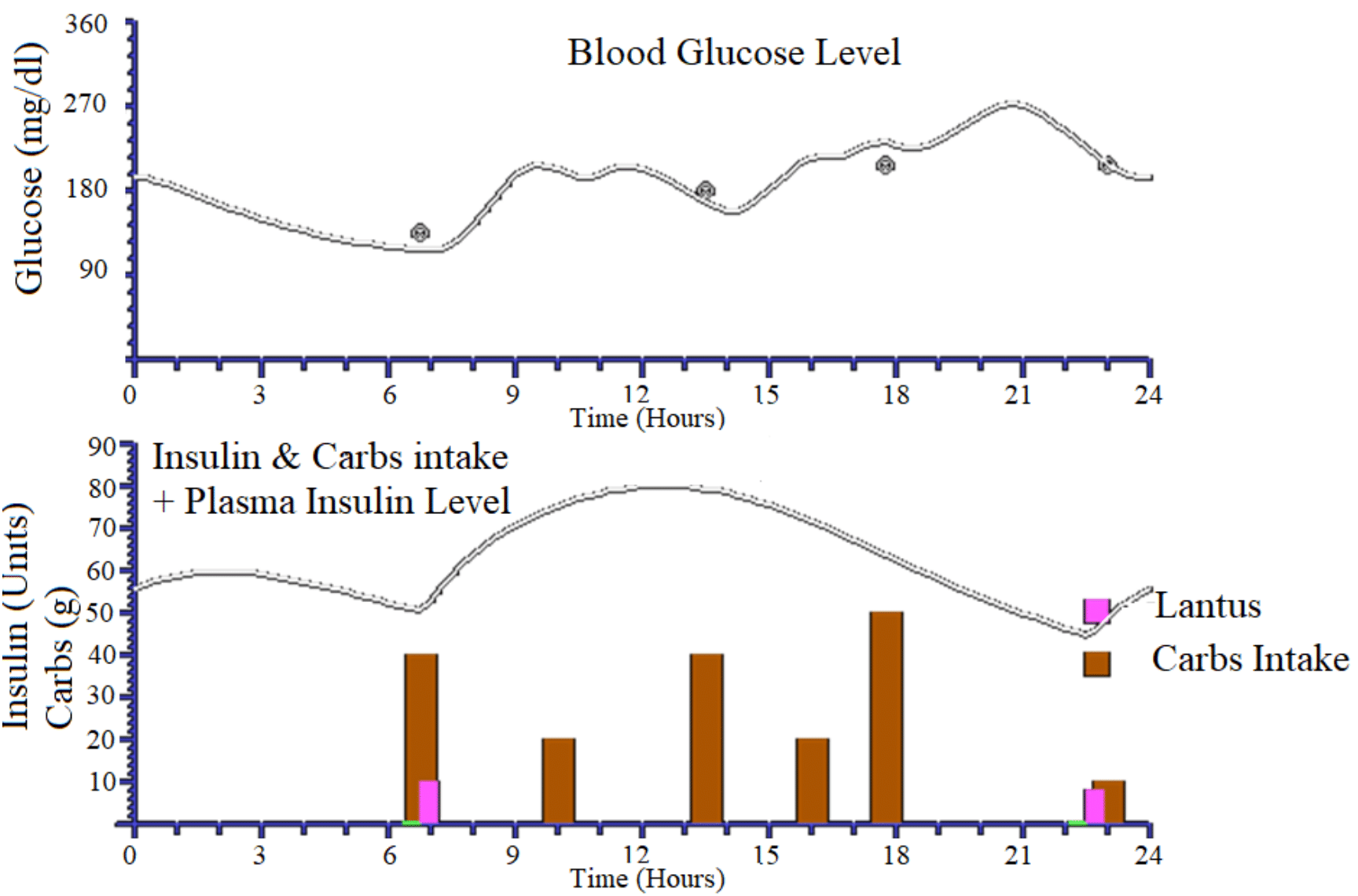}}\label{}
	\caption{Analysis of glucose fluctuations with respect to given insulin doses and carbs intake of 5 consecutive cases} 
	\label{sim}
\end{figure}

For glucose level balancing point of view, experimental analysis is required to validate proposed paradigm. To perform the experimental analysis, 5 different cases have been taken from 70 samples (Hyperglycemia). 70 samples are listed out for bifurcation of samples, who were taking insulin treatment to maintain normal glucose level. 10 samples were found with insulin level treatment out of 70 samples. Out of 10 cases, 5 cases have been chosen for keen observation to validate the proposed system. These 5 cases were prescribed with initial scheduled insulin and diet plans as per the expert medical consultant. The carbs intake (food), insulin doses and glucose levels are analyzed for 24 hrs. These values are required to prepare the data by decision model, which would have model for finding auxiliary parameters. The data is needed to prepare further treatment by medical consultant with less effort of time and stages of treatment. This would be helpful to patient to maintain the glucose values at normal range.
5 cases are represented with different conditions of blood glucose fluctuations and insulin doses at specific time as prescribed. Glucose levels were observed according to provided different insulin doses and carbs intake. The analysis is represented in Fig \ref{sim}.
The simulated results represented the parameters to provide further treatment to maintain the normal glucose levels. The comparison table represented the statistical parameters and technology specifications of proposed work compared to prior related work. The tabular representation shows the advancement in current work, which is represented in Table \ref{table_example}.

\begin{table*}[htbp]
	\caption{Comparison with Non-invasive Works}
	\label{table_example}
	\centering
	\begin{tabular}{llllll}
		\hline
		\textbf{Works}&{$R^{2}$}&MARD&Technology&Measurement&Application\\
		\textbf{}&value&(\%)&Specification&Range (mg/dl)&\\
		\hline		 \hline
		Jain, et al.	\cite{Jain_IEEE-MCE_2020-Jan_iGLU1}&0.90&9.96 &mNIRS&80-250&Glucose\\
		&&&&&Measurement\\
		\hline
		Singh, et al.	\cite{newsingh8727488}&0.8&-&Saliva&90-150&Glucose\\
		&&&&&Measurement\\
		\hline
		Joshi, et al. \cite{iGLU3}&0.96&12.6&mNIRS&80-300&Glucose-Insulin\\
		&&&&&Model\\
		\hline
		Murad, et al.\cite{new9431682}&-&-&Simulation&-&Glucose Detection\\
		\hline
		Kirubakaran, et al.\cite{newkirubakaran2023antiallergic}&0.91&-&Microwave&110-400&Glucose\\
		&&&Active Sensor&&Measurement\\
		\hline
		Mohammadi, et al.\cite{newmohammadi2023dual}&-&-&Microwave&-&Glucose\\
		&&&Resonator&&Detection\\
		\hline
		\textbf{Proposed Work}&\textbf{0.97}&\textbf{12.5}&\textbf{mNIRS with SBP}&\textbf{80-400}&\textbf{Glucose Balancing}\\
		\textbf{iGLU 4.0}&\textbf{}&\textbf{}&\textbf{+Body Temperature}&&\textbf{Paradigm}\\
		\hline
	\end{tabular}
\end{table*}

\section{Conclusion and Future Direction}

The proposed paradigm of glucose balancing consists of precise glucose measurement along with estimation of physiological parameters. The proposed system provides the feature to generate supporting data, which would be helpful to the expert for prior level of treatment (prescribed treatment without conventional testings). A multi-wave spectroscopy technique with unique combination of Physiological parameters based non-invasive blood glucose monitoring system is highlighted. The proposed system has an advantage of collecting data without prior setting of the prototype system for precise detection. The proposed system is data demonstrating system to provide better treatment with less efforts, time and source usage. Hence, the proposed system will be comparatively cost effective solution. The analysis of results has been done in real time. All samples are collected from healthy, hypoglycemic and hyperglycemic patients with standard protocol of sample collection. The physiological parameters are also considered at the time of sample collection for system validation. During statistical analysis using proposed computation models, 0.97 coefficient of determination is calculated using proposed deep neural network based model. After analysis, 94\% samples have been found in the desire zone. For the advancement of proposed work, some more physiological parameters are required to involve for precision point of view. Most of blood parameters need to examine along with glucose level values for multi vitals monitoring. 

In the future research, we will design a non-invasive system for glycated hemoglobin (HbA1c) test. This blood glucose test indicates your average blood glucose level for the past two to three months. HbA1c test is always considered preliminary test for type 2 diabetic patients. This system has to be propose with features of Internet-of-Medical-Things (IoMT). By this feature with HbA1c test, the system will be cost effective and rapid measurement solution. 

\section*{Acknowledgment}
The authors would like to express their sincere gratitude to Dispensary, Malaviya National of Technology and System Level Design and Calibration Testing Lab. There is special thanks to Nirma University to encourage for continuing research with supports. The research work was funded by institute fellowship from Ministry of HRD, Govt. of India to pursue the Ph.D work at MNIT Jaipur.

\newpage
\bibliographystyle{IEEEtran}
\bibliography{arXiv_2021_iGLU4_NIR-Body-Temperature}


\section*{Authors' Biographies}
\begin{minipage}[htbp]{\columnwidth}
	\begin{wrapfigure}{l}{1.0in}
		\vspace{-0.4cm}
		\includegraphics[width=1.0in,keepaspectratio]{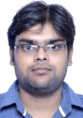}
		\vspace{-0.3cm}
	\end{wrapfigure}
	\noindent
	\textbf{Prateek Jain} (member IEEE) earned his B.E. degree in Electronics Engineering from Jiwaji University, India in 2010 and Master degree from ITM University Gwalior. He obtained PhD from Malaviya National Institute of Technology, Jaipur. He was awarded from MHRD fellowship during 2016-2020. He is currently Assistant Professor in Electronics \& Instrumentation Engg. Deptt., Institute of Technology, Nirma University, Ahmedabad (India). He was senior assistant professor in school of electronics engg. (SENSE), VIT AP University, AP since June 2020-Dec.2023. His current research interest includes Real-time system design, Biomedical Instrumentation, Low power VLSI design and Biomedical Systems. He is an author of 25 peer-reviewed publications. He is a regular reviewer of journals and 10 conferences. He was resource person in reputed universities for technical programs.
\end{minipage}

\vspace{0.8cm}

\begin{minipage}[htbp]{\columnwidth}
	\begin{wrapfigure}{l}{1.0in}
		\vspace{-0.4cm}
		\includegraphics[width=1.0in,keepaspectratio]{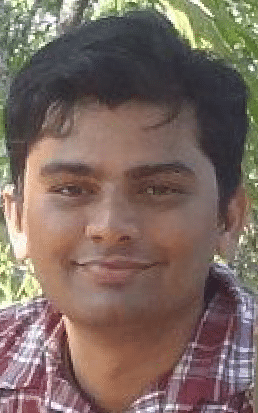}
		\vspace{-0.5cm}
	\end{wrapfigure}
	\noindent
	\textbf{Amit M. Joshi} (M'08) has completed his M.Tech (by research) in 2009 and obtained Doctoral of Philosophy degree (Ph.D) from National Institute of Technology, Surat in August 2015. He is currently an Assistant Professor at National Institute of Technology, Jaipur since July 2013. His area of specialization is Biomedical signal processing, Smart healthcare, VLSI DSP Systems and embedded system design. He has published six book chapters and also published 50+ research articles in peer reviewed international journals/conferences. He has served as a reviewer of technical journals such as IEEE Transactions, Springer, Elsevier and also served as Technical Programme Committee member for IEEE conferences. He also received UGC Travel fellowship, SERB DST Travel grant  and CSIR Travel fellowship to attend IEEE Conferences in VLSI and Embedded System. He has served session chair at various IEEE Conferences like TENCON -2016, iSES-2018, ICCIC-14. He has already supervised 18 M.Tech projects and 14 B.Tech projects in the field of VLSI and Embedded Systems and VLSI DSP systems. He is currently supervising six  Ph.D. students.
\end{minipage}

\vspace{0.5cm}

\begin{minipage}[htbp]{\columnwidth}
	\begin{wrapfigure}{l}{1.00in}
		\vspace{-0.3cm}
		\includegraphics[width=1.0in,keepaspectratio]{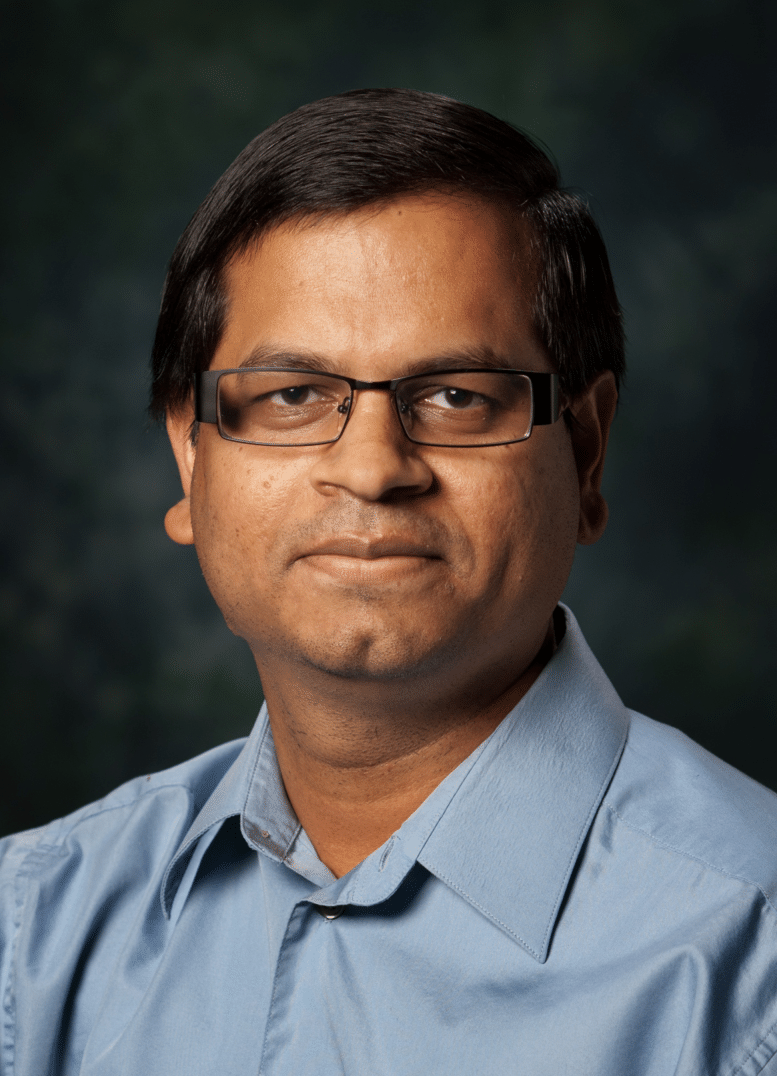}
		\vspace{-0.5cm}
	\end{wrapfigure}
	\noindent
	\textbf{Saraju P. Mohanty} (Senior Member, IEEE) received the bachelor’s degree (Honors) in electrical engineering from the Orissa University of Agriculture and Technology, Bhubaneswar, in 1995, the master’s degree in Systems Science and Automation from the Indian Institute of Science, Bengaluru, in 1999, and the Ph.D. degree in Computer Science and Engineering from the University of South Florida, Tampa, in 2003. He is a Professor with the University of North Texas. His research is in ``Smart Electronic Systems’’ which has been funded by National Science Foundations (NSF), Semiconductor Research Corporation (SRC), U.S. Air Force, IUSSTF, and Mission Innovation. He has authored 450 research articles, 5 books, and 10 granted and pending patents. His Google Scholar h-index is 53 and i10-index is 225 with 12,000 citations. He is regarded as a visionary researcher on Smart Cities technology in which his research deals with security and energy aware, and AI/ML-integrated smart components. He introduced the Secure Digital Camera (SDC) in 2004 with built-in security features designed using Hardware Assisted Security (HAS) or Security by Design (SbD) principle. He is widely credited as the designer for the first digital watermarking chip in 2004 and first the low-power digital watermarking chip in 2006. He is a recipient of 16 best paper awards, Fulbright Specialist Award in 2020, IEEE Consumer Electronics Society Outstanding Service Award in 2020, the IEEE-CS-TCVLSI Distinguished Leadership Award in 2018, and the PROSE Award for Best Textbook in Physical Sciences and Mathematics category in 2016. He has delivered 21 keynotes and served on 14 panels at various International Conferences. He has been serving on the editorial board of several peer-reviewed international transactions/journals, including IEEE Transactions on Big Data (TBD), IEEE Transactions on Computer-Aided Design of Integrated Circuits and Systems (TCAD), IEEE Transactions on Consumer Electronics (TCE), and ACM Journal on Emerging Technologies in Computing Systems (JETC). He has been the Editor-in-Chief (EiC) of the IEEE Consumer Electronics Magazine (MCE) during 2016-2021. He served as the Chair of Technical Committee on Very Large Scale Integration (TCVLSI), IEEE Computer Society (IEEE-CS) during 2014-2018 and on the Board of Governors of the IEEE Consumer Electronics Society during 2019-2021. He serves on the steering, organizing, and program committees of several international conferences. He is the steering committee chair/vice-chair for the IEEE International Symposium on Smart Electronic Systems (IEEE-iSES), the IEEE-CS Symposium on VLSI (ISVLSI), and the OITS International Conference on Information Technology (OCIT). He has mentored 3 post-doctoral researchers, and supervised 15 Ph.D. dissertations, 26 M.S. theses, and 18 undergraduate projects.

\end{minipage}

\end{document}